\documentclass[12pt]{article}
\usepackage[dvips]{color}
\usepackage{epsfig}
\usepackage{amsmath}
\usepackage{graphicx}
\def\Box{\hbox{$\rlap{$\sqcup$}\sqcap$}}
\begin{document}
\title{\bf Parametrizations of Dark Energy Models in the Background of General
Non-canonical Scalar Field in $D$-dimensional Fractal Universe }

\author{\bf{Ujjal Debnath}
\thanks{ujjaldebnath@gmail.com} \\
Department of Mathematics, Indian Institute of Engineering\\
Science and Technology, Shibpur, Howrah-711 103, India.\\\\
\bf{Kazuharu Bamba}
\thanks{bamba@sss.fukushima-u.ac.jp }\\
Division of Human Support System, Faculty of Symbiotic Systems
Science,\\ Fukushima University, Fukushima 960-1296, Japan.}

\maketitle

\begin{abstract}
We have explored non-canonical scalar field model in the
background of non-flat $D$-dimensional fractal Universe on the
condition that the matter and scalar field are separately
conserved. The potential $V$, scalar field $\phi$, function $f$,
densities, Hubble parameter and deceleration parameter can be
expressed in terms of the redshift $z$ and these depend on the
equation of state parameter $w_{\phi}$. We have also investigated
the cosmological analysis of four kinds of well known
parametrization models. In graphically, we have analyzed the
natures of potential, scalar field, function $f$, densities, the
Hubble parameter and deceleration parameter. As a result, the best
fitted values of the unknown parameters ($w_{0},w_{1}$) of the
parametrization models due to the joint data analysis
(SNIa+BAO+CMB+Hubble) have been found. Furthermore, the minimum
values of $\chi^{2}$ function have been obtained. Also we have
plotted the graphs for different confidence levels 66\%, 90\% and
99\% contours for ($w_{0},~w_{1}$) by fixing the other parameters.
\end{abstract}

\section{Introduction}

During last two decades, several observations like type Ia
Supernovae, Cosmic Microwave Background (CMB) radiations, large
scale structure (LSS), Sloan Digital Sky Survey (SDSS), Wilkinson
Microwave Anisotropy Probe (WMAP), Planck observations
\cite{Perlmutter,Riess,Briddle,Spergel,Lidd,Kin,Sper2,Kom,Cala,Wang0,Zhao0,Ade0}
suggest that our Universe is experiencing an accelerated expansion
due to some unknown exotic fluid which generates sufficient
negative pressure, known as dark energy (DE). Although a long-time
debate has been done on this well-reputed and interesting issue of
modern cosmology, we still have little knowledge about DE. The
most appealing and simplest candidate for DE is the cosmological
constant $\Lambda$. Since the source of the DE still now unknown,
so several candidates of the DE models have been proposed in the
literatures where the scalar field plays a significant role in
cosmology as they are sufficiently complicated to produce the
observed dynamics. Recently many cosmological models have been
constructed by introducing dark energies such as quintessence,
Phantom, Tachyon, k-essence, dilaton, Hessence, DBI-essence, ghost
condensate, quintom, Chaplygin gas models, interacting dark energy
models
\cite{Peebles,Cald,Cald1,Akrami1,Sen,Arme,gas,Wei,Gum,Mart,Ark,Piaz,Feng,Guo,Kamen,Amen,Zhang,Bamba00}.
A review on dynamics of dark energy has been studied in ref
\cite{Sami}. Another approach to explore the accelerated expansion
of the universe is the modified theories of gravity. The main
candidates of modified gravity includes brane world models, DGP
brane, LQC, Brans-Dicke gravity, Gauss-Bonnet gravity,
Horava-Lifshitz gravity, $f(R)$ gravity, $f(T)$ gravity, $f(G)$
gravity etc
\cite{Sahni,Dvali,Brans,An,Ab,Hora,Lin,Yer,Noj,An1,Eis,Cog,Mar}.
Also for recent reviews on the issue of dark energy and modified
gravity theories, see, for instance,~\cite{Bamba,Nojiri,
Capozziello, Nojiri1, Capozziello1,Cai0,Bamba1}.\\

Motivated by high energy physics, the scalar field models play an
significant role to explain the nature of DE due to its simple
dynamics \cite{Copl,Tsu}. In the very early epoch of the Universe,
it is strongly believed that the universe had a very rapid
exponential growth during a very short era (which is known as
inflation). During this era, there was a single canonical scalar
field called {\it inflaton} \cite{Lidd} which has a canonical
kinetic energy term in the Lagrangian density. The simplest form
of the canonical scalar field is known as quintessence field,
which alone cannot explain the phantom crossing. Also the
canonical scalar field cannot fully explain several cosmological
natures of the Universe. So non-canonical scalar field was
proposed to solve several important cosmological problems
\cite{Mukh,Fang}. For instance, the non-canonical scalar field can
resolve the coincidence problem \cite{Lee}. So it is reasonable to
consider the non-canonical scalar field as a viable cosmological
model of DE candidate. In general, the non-canonical scalar field
involves higher order derivatives term of a scalar field in the
Horndeski Lagrangian \cite{Horn}. For instance, the $k$-essence is
one of the simple form of a non-canonical scalar field model. In
the present work, we will consider non-canonical scalar field
model with general form of $k$-essence Lagrangian
\cite{Melchi,Unni1,Unni2,Das11,Mamon0,Mamon1,Jibitesh}.\\

The idea of fractal effects in the Einstein's equations is another
approach to cosmic acceleration in a gravity theory. Calcagni
\cite{1GC10,Cal2} discussed the quantum gravity phenomena and
different cosmological properties in fractal universe. Fractal
features of quantum gravity and Cosmology in $D$-dimensions have
also been investigated. Also it has been discussed the properties
of a scalar field model at classical and quantum level.
The Multi-scale gravity and cosmology have been studied in
\cite{Calg00}. Karami et al \cite{Karami} have investigated the
holographic, new agegraphic and ghost dark energy models in the
framework of fractal cosmology. Lemets et al \cite{Lemets} have
studied the interacting dark energy models in the fractal universe
with the interaction between dark energy and dark matter. Sheykhi
et al \cite{Sheykhi} have analyzed the thermodynamical properties
on the apparent horizon in the fractal universe. Chattopadhyay et
al \cite{Chatto} have discussed some special forms of holographic
Ricci dark energy in fractal Universe. Halder et al \cite{Halder}
have presented a comparative study of different entropies in
fractal Universe. Maity and Debnath \cite{Maity} have studied the
co-existence of modified Chaplygin gas and other dark energies in
the framework of fractal Universe. Jawad et al \cite{Jawad} have
studied the implications of pilgrim dark energy models in the
fractal Universe. Das et al \cite{Das} have described cosmic
scenario in the framework of fractal Universe. Sardi et al
\cite{Sardi} have studied an
interacting new holographic dark energy in the framework of fractal cosmology.\\

In the present work, we consider the non-canonical scalar field
model in the background of effective fractal spacetime. We shall
extend the work of ref \cite{Mamon1} in the framework of
$D$-dimensional fractal spacetime by considering more general
power law form of kinetic term. We shall try to obtain the
observationally viable model to describe the nature of the
equation of state (EoS) parameter $w_{\phi}(z)$ for the scalar
field and the deceleration parameter $q(z)$ which are functions of
redshift $z$. To describe the DE evolutions, we will choose four
kinds of well known parametrizations models for different choices
of $w_{\phi}(z)$. We found the best fit values and confidence
contours of two parameters for different forms of $w_{\phi}$ by
considering the observational data analysis of SNIa, BAO, CMB and
Hubble. The paper is organized as follows: In section 2, we
discuss a general non-canonical scalar field model in the
framework of non-flat model of $D$-dimensional fractal Universe.
By suitable choice of the function $f(\phi)$, we obtain potential
function $V(\phi)$ and Hubble parameter in terms of $z$ and EoS
parameter $w_{\phi}(z)$. Then we choose four kinds of EoS
parameter $w_{\phi}(z)$ to obtain $H(z)$. In section 3, we provide
the joint data analysis mechanism for the observations of SNIa,
BAO, CMB and Hubble. Finally we discuss the results of the work in
section 4.

\section{Non-Canonical Scalar Field Model in Fractal Universe}

In the {\it fractal Universe} \cite{1GC10}, the space and time
co-ordinates scale are satisfying $[x^{\mu}]=-1,~\mu=0,1,...,
D-1$, where $D$ is the topological dimension of the embedding
space-time. Also in the action for the fractal Universe
\cite{1GC10,Das}, the standard measure can be replaced by a
non-trivial measure which appears in Lebesgue-Stieltjes integrals:
$d^{D}x \rightarrow d\varrho(x)$ with
$[\varrho]=-D\alpha,~\alpha\ne 1$, where $\alpha$ describes the
fraction of states. Here the measure is considered as general
Borel $\varrho$ on a fractal set. So in $D$ dimensions, $(M,
\varrho)$ denotes the metric space-time where $M$ is equipped with
measure $\varrho$. Here the probability measure $\varrho$ is a
continuous function with $d\varrho(x)=(d^{D}x) v(x)$, which is the
{\it Lebesgue-Stieltjes measure} and $v$ is known as the weight
function or {\it fractal function}.\\

In the Einstein's gravity, the total action for the scalar field
model in effective fractal space-time is given by \cite{1GC10}
\begin{equation}
S=S_{g}+S_{sf}+S_{m}
\end{equation}
where the $ansatz$ for the gravitational action is
\begin{equation}
S_{g}=\frac{1}{2\kappa^{2}}\int
 d\varrho(x)\sqrt{-g}~(R-\omega\partial_{\mu}v\partial^{\mu}v)~,
\end{equation}
the scalar field action is given by
\begin{equation}
S_{sf}=\int d\varrho(x)\sqrt{-g}~{\cal
 L}(\phi,X)
\end{equation}
and the matter action is given by
\begin{equation}
S_{m}=\int d\varrho(x)\sqrt{-g}~{\cal L}_{m}
\end{equation}
Here $g$ is the determinant of the dimensionless metric
$g_{\mu\nu}$, $\kappa^{2}=8\pi G$ is Newton's constant, $R$ is the
Ricci scalar and the term proportional to $\omega$ (fractal
parameter) has been added because $v$ (fractal function), like the
other geometric quantity $g_{\mu\nu}$, is dynamical. Also ${\cal
L}(\phi,X)$ is the Lagrangian density for scalar field $\phi$,
$X(=\frac{1}{2}\partial_{\mu}\phi\partial^{\mu}\phi)$ is the
kinetic term, ${\cal L}_{m}$ is the matter Lagrangian.\\

In the general form of non-canonical scalar field model, the
Lagrangian density can be expressed as \cite{Melchi,Mamon1}
\begin{equation}
{\cal L}(\phi,X)=f(\phi)X\left(\kappa^{4}X \right)^{n-1}-V(\phi)
\end{equation}
where $f(\phi)$ is the arbitrary function and $V(\phi)$ is the
corresponding potential for the non-canonical scalar field $\phi$.
For $n=1$ and $f(\phi)=1$, it reduces to usual canonical scalar
field Lagrangian. For $n=1$ and $f(\phi)=-1$, we get the phantom
scalar field model. For $f(\phi)=1$, it reduces to
particular form of general non-canonical scalar field model \cite{Fang}.\\

The expressions for the energy density $\rho_{\phi}$ and pressure
$p_{\phi}$ associated with the non-canonical scalar field $\phi$
are given by
\begin{equation}\label{6}
\rho_{\phi}=2X~\frac{\partial {\cal L}}{\partial X}-{\cal
L}=(2n-1)2^{-n}\kappa^{4(n-1)}f(\phi)\dot{\phi}^{2n}+V(\phi)
\end{equation}
\begin{equation}\label{7}
p_{\phi}={\cal
L}=2^{-n}\kappa^{4(n-1)}f(\phi)\dot{\phi}^{2n}-V(\phi)
\end{equation}

Now we consider the line-element for $D$-dimensional non-flat
Friedmann-Robertson-Walker (FRW) universe as
\begin{equation}
ds^{2}=-dt^{2}+a^{2}(t)\left[\frac{dr^{2}}{1-kr^{2}}+r^{2}d\Omega^{2}_{D-2}
\right]
\end{equation}
where $a(t)$ is the scale factor and $k~(=0,\pm 1)$ is the
curvature scalar.\\

Taking the variation of the action given in (1) with respect to
the $D$-dimensional FRW metric $g_{\mu\nu}$, we obtain the
Friedmann equations in a fractal universe as \cite{1GC10,Maity}
\begin{equation}\label{9}
\left(\frac{D}{2}-1\right)H^{2}+\frac{k}{a^{2}}+H\frac{\dot{v}}{v}-\frac{1}{2}\frac{\omega}{D-1}\dot{v}^{2}
=\frac{\kappa^{2}}{D-1}~\rho
\end{equation}
and
\begin{equation}\label{10}
(D-2)\left(\dot{H}+H^{2}-H\frac{\dot{v}}{v}+\frac{\omega}{D-1}\dot{v}^{2}
\right)-\frac{\Box v}
{v}=-\frac{\kappa^{2}}{D-1}~[(D-3)\rho+(D-1)p]
\end{equation}
where $H(=\frac{\dot{a}}{a})$ is the Hubble parameter and $\Box v$
(where $\Box$ is the D'Alembertian operator) is defined by
\begin{equation}
\Box
v=\frac{1}{\sqrt{-g}}~\partial^{\mu}(\sqrt{-g}~\partial_{\mu}v)
\end{equation}
which can be simplified to the following form:
\begin{equation}\label{12}
\Box v=-[\ddot{v}+(D-1)H\dot{v}]
\end{equation}

Here the total energy density is $\rho=\rho_{m}+\rho_{\phi}$ and
total pressure is $p=p_{m}+p_{\phi}$ where $\rho_{m}$ and $p_{m}$
are respectively energy density and pressure for matter. The
continuity equation for fractal Universe is given by
\begin{equation}\label{13}
\dot{\rho}+\left[(D-1)H+\frac{\dot{v}}{v}\right](\rho+p)=0
\end{equation}
Since the fractal function $v$ is time dependent, so we can choose
the power law form of of $v$ in terms of the scale factor as
$v=v_{0}a^{\beta}$ where $v_{0}$ and $\beta$ are positive
constants \cite{Sheykhi,Chatto,Halder,Maity,Jawad,Das,Sardi}. The
parameter $\beta$ is related to the Hausdorff dimension of the
physical space-time. So the equations (\ref{9}), (\ref{10}) and
(\ref{13}) reduce to

\begin{equation}\label{14}
\left[\frac{D}{2}-1+\beta-\frac{\omega
v_{0}^{2}\beta^{2}a^{2\beta}}{2(D-1)}\right]H^{2}+\frac{k}{a^{2}}
=\frac{\kappa^{2}}{D-1}~(\rho_{m}+\rho_{\phi})~,
\end{equation}
\begin{eqnarray}\label{15}
(D-2+\beta)\dot{H}+\left[(D-2)\left(1-\beta+\frac{\omega
v_{0}^{2}\beta^{2}a^{2\beta}}{D-1} \right)+\beta^{2}+(D-1)\beta
\right]H^{2} \nonumber\\
=-\frac{\kappa^{2}}{D-1}~[(D-3)(\rho_{m}+\rho_{\phi})+(D-1)(p_{m}+p_{\phi})]
\end{eqnarray}
and
\begin{equation}\label{16}
(\dot{\rho}_{m}+\dot{\rho}_{\phi})+(\beta+D-1)H(\rho_{m}+\rho_{\phi}+p_{m}+p_{\phi})=0
\end{equation}

Now we assume that the matter and scalar field are separately
conserved. So the conservation equations for matter and scalar
field are given by
\begin{equation}\label{17}
\dot{\rho}_{m}+(\beta+D-1)H(\rho_{m}+p_{m})=0
\end{equation}
and
\begin{equation}\label{18}
\dot{\rho}_{\phi}+(\beta+D-1)H(\rho_{\phi}+p_{\phi})=0
\end{equation}
Solving equation (\ref{17}), we get the expression of energy
density of matter as
\begin{equation}
\rho_{m}(z)=\rho_{m0}(1+z)^{(\beta+D-1)(1+w_{m})}
\end{equation}
where $\rho_{m0}$ is the present value of the energy density,
$w_{m}=\frac{p_{m}}{\rho_{m}}$ is the constant equation of state
parameter for matter and $z~(=\frac{1}{a}-1)$ is the
redshift (choosing $a_{0}=1$).\\

From equations (\ref{6}) and (\ref{7}), we obtain the the
potential $V(\phi)$ and $f(\phi)$ as in the following forms:
\begin{equation}\label{20}
V(\phi)=\frac{1}{2n}[1-(2n-1)w_{\phi}]\rho_{\phi}
\end{equation}
and
\begin{equation}\label{21}
f(\phi)\dot{\phi}^{2n}=\frac{2^{n}}{2n\kappa^{4(n-1)}}(1+w_{\phi})\rho_{\phi}
\end{equation}
Now we obtain the equation of state parameter $w_{\phi}$ for
scalar field as
\begin{equation}
w_{\phi}=\frac{p_{\phi}}{\rho_{\phi}}=\frac{\kappa^{4(n-1)}f(\phi)\dot{\phi}^{2n}
-2^{n}V(\phi)}{(2n-1)\kappa^{4(n-1)}f(\phi)\dot{\phi}^{2n}+2^{n}V(\phi)}
\end{equation}
Now integrating equation (\ref{18}), we obtain \cite{Mamon1}
\begin{equation}
\rho_{\phi}(z)=\rho_{\phi
0}~exp\left((\beta+D-1)\int_{0}^{z}\frac{1+w_{\phi}(z)}{1+z}~dz
\right)
\end{equation}
where $\rho_{\phi 0}$ is the present value of energy density for
scalar field. To solve the scalar field $\phi$, we assume
\cite{Mamon1}
\begin{equation}
f(z)=\left(\frac{f_{0}}{H}\right)^{2n}
\end{equation}
where $f_{0}$ is constant. In ref \cite{Mamon1}, the authors have
assumed $n=2$ and studied the features of the model. From equation
(\ref{21}), we obtain
\begin{equation}
\phi(z)=\phi_{0}+\frac{\sqrt{2}}{f_{0}(2n)^{\frac{1}{2n}}\kappa^{\frac{2(n-1)}{n}}}\int_{0}^{z}
\frac{(1+w_{\phi}(z))^{\frac{1}{2n}}\rho_{\phi}(z)^{\frac{1}{2n}}}{1+z}~dz
\end{equation}
where $\phi_{0}$ is the present value of $\phi$. From equation
(\ref{20}), we obtain the potential function as
\begin{equation}
V(z)=\frac{\rho_{\phi
0}}{2n}[1-(2n-1)w_{\phi}(z)]~exp\left((\beta+D-1)\int_{0}^{z}\frac{1+w_{\phi}(z)}{1+z}~dz
\right)
\end{equation}

From equation (\ref{14}), we obtain \cite{Mamon1}
\begin{equation}\label{27}
H^{2}(z)=\frac{\xi
H_{0}^{2}\left[\Omega_{m0}(1+z)^{(\beta+D-1)(1+w_{m})}+\Omega_{\phi
0}~exp\left((\beta+D-1)\int_{0}^{z}\frac{1+w_{\phi}(z)}{1+z}~dz
\right)-\Omega_{k0}(1+z)^{2}\right]}{\left[\xi+\omega
v_{0}^{2}\beta^{2}-\omega
v_{0}^{2}\beta^{2}(1+z)^{-2\beta}\right]}
\end{equation}
where $H_{0}$ is the present value of the Hubble parameter,
$\xi=(D-1)(D-2+2\beta)-\omega v_{0}^{2}\beta^{2}$,
$\Omega_{m0}=\frac{2\kappa^{2}\rho_{m0}}{\xi H_{0}^{2}}$ and
$\Omega_{\phi 0}=\frac{2\kappa^{2}\rho_{\phi 0}}{\xi H_{0}^{2}}$
and $\Omega_{k0}=\frac{2(D-1)k}{\xi H_{0}^{2}}$ are present value
of the density parameters of matter, scalar field and curvature
scalar respectively satisfying $\Omega_{m0}+\Omega_{\phi
0}-\Omega_{k0}=1$. The deceleration parameter $q(z)$ can be
written as
\begin{equation}
q(z)=-\frac{\ddot{a}}{aH^{2}}=-1+\frac{(1+z)}{2H^{2}}~\frac{dH^{2}}{dz}
\end{equation}
Since expression of $H(z)$ is given in equation (\ref{27}), so
$q(z)$ can be expressed in term of $z$ analytically. Now we see
that the functions
$\rho_{\phi}(z),~\phi(z),~V(z),f(z),~H(z),~q(z)$ completely depend
on the EoS parameter function $w_{\phi}(z)$ with number of
constant parameters. In the next subsections, we will consider
different well known parameterization forms of $w_{\phi}(z)$ and
investigate the natures of Hubble parameter, deceleration
parameter, scalar field and its potential in different models.

\subsection{\normalsize\bf{Model I : Linear Parameterization}}

The equation of state parameter for linear parametrization is
\cite{Coor} $w_{\phi}(z)=w_0+{w_1}z$, where $w_0$ and $w_1$ are
constants in which $w_{0}$ represents the present value of
$w_{\phi}(z)$. The energy density of the model then gives rise to

\begin{equation}
\rho_{\phi}(z)=\rho_{\phi
0}(1+z)^{(\beta+D-1)(1+w_0-{w_1})}e^{(\beta+D-1){w_1}z}
\end{equation}

From equation (\ref{27}), the Hubble parameter can be expressed as

\begin{equation}\label{30}
H^{2}(z)=\frac{\xi
H_{0}^{2}\left[\Omega_{m0}(1+z)^{(\beta+D-1)(1+w_{m})}+\Omega_{\phi
0}~(1+z)^{(\beta+D-1)(1+w_0-{w_1})}e^{(\beta+D-1){w_1}z}-\Omega_{k0}(1+z)^{2}\right]}{\left[\xi+\omega
v_{0}^{2}\beta^{2}-\omega
v_{0}^{2}\beta^{2}(1+z)^{-2\beta}\right]}
\end{equation}

\begin{figure}
\includegraphics[height=2.5in]{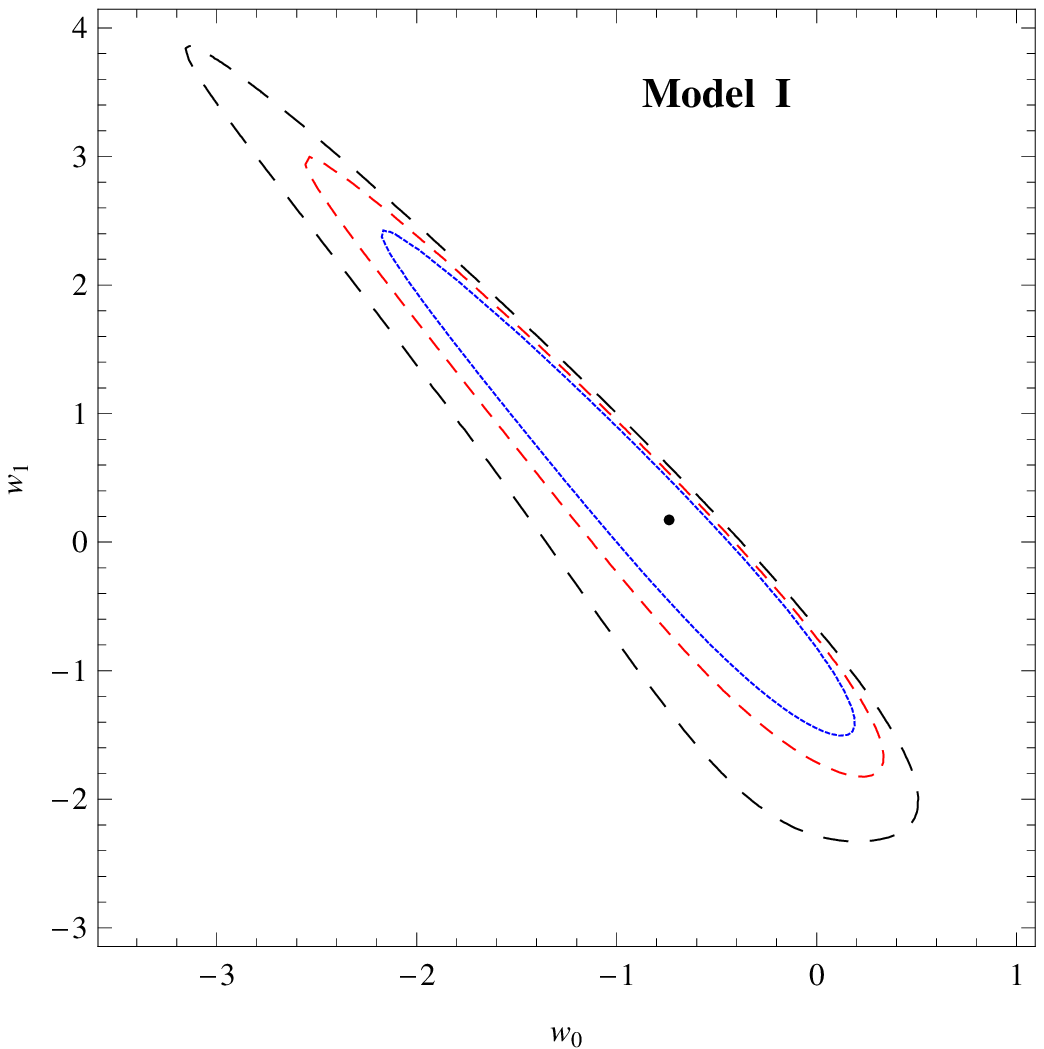}~~~~
\includegraphics[height=2.0in]{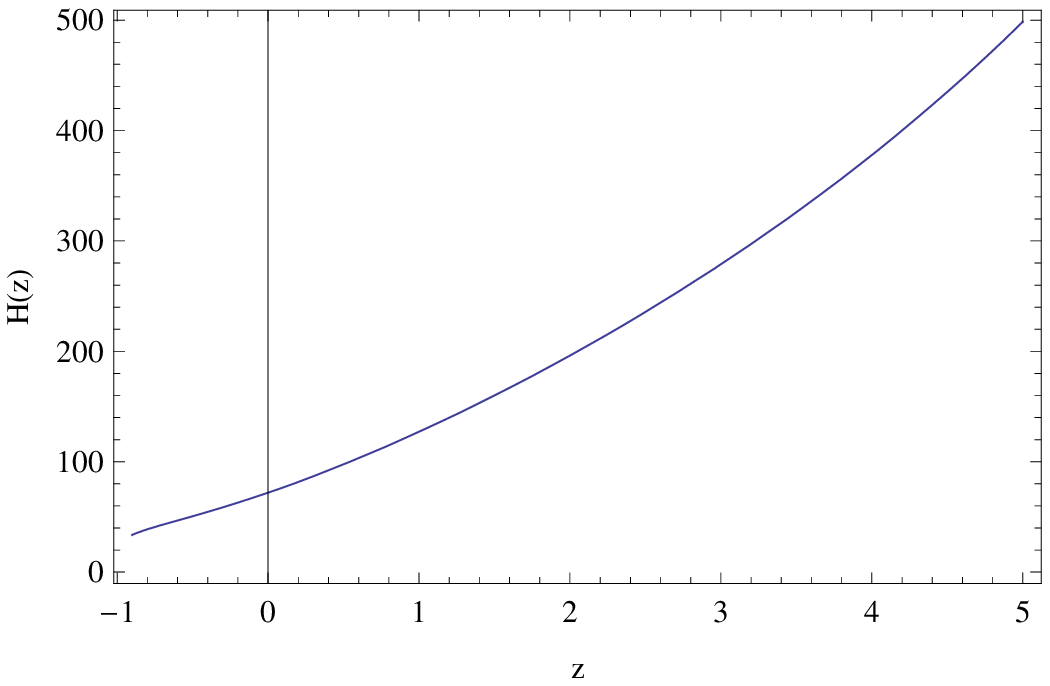}\\

~~~~~~~~~~~~~~~~~~~~~~~~~~~~~~~~~~Fig.1~~~~~~~~~~~~~~~~~~~~~~~~~~~~~~~~~~~~~~~~~~~~~~~~~~~~~~~~~~~~~~~~Fig.2\\
\vspace{2mm}

\includegraphics[height=2.0in]{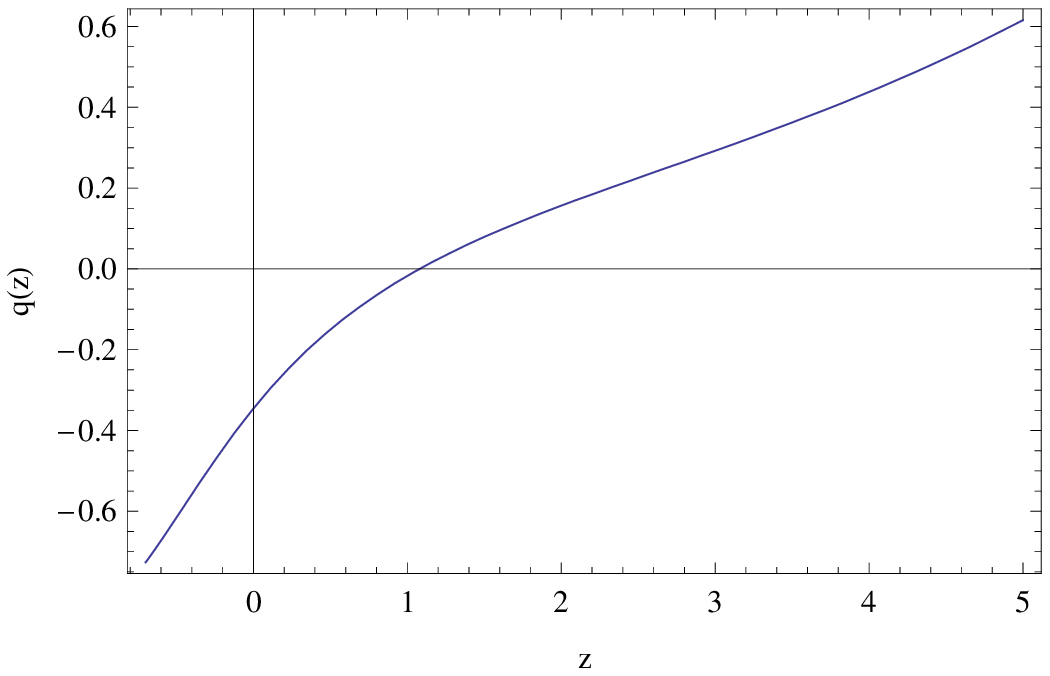}~~~~
\includegraphics[height=2.0in]{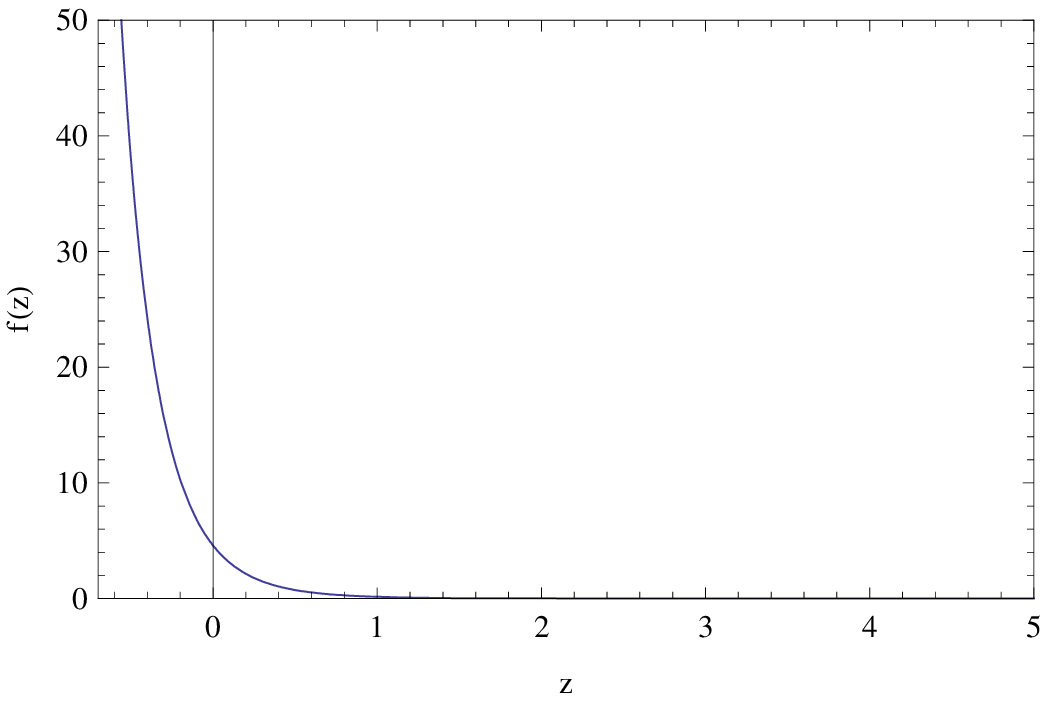}\\

~~~~~~~~~~~~~~~~~~~~~~~~~~~~~~~~~~Fig.3~~~~~~~~~~~~~~~~~~~~~~~~~~~~~~~~~~~~~~~~~~~~~~~~~~~~~~~~~~~~~~~~Fig.4\\
\vspace{2mm}

\includegraphics[height=2.0in]{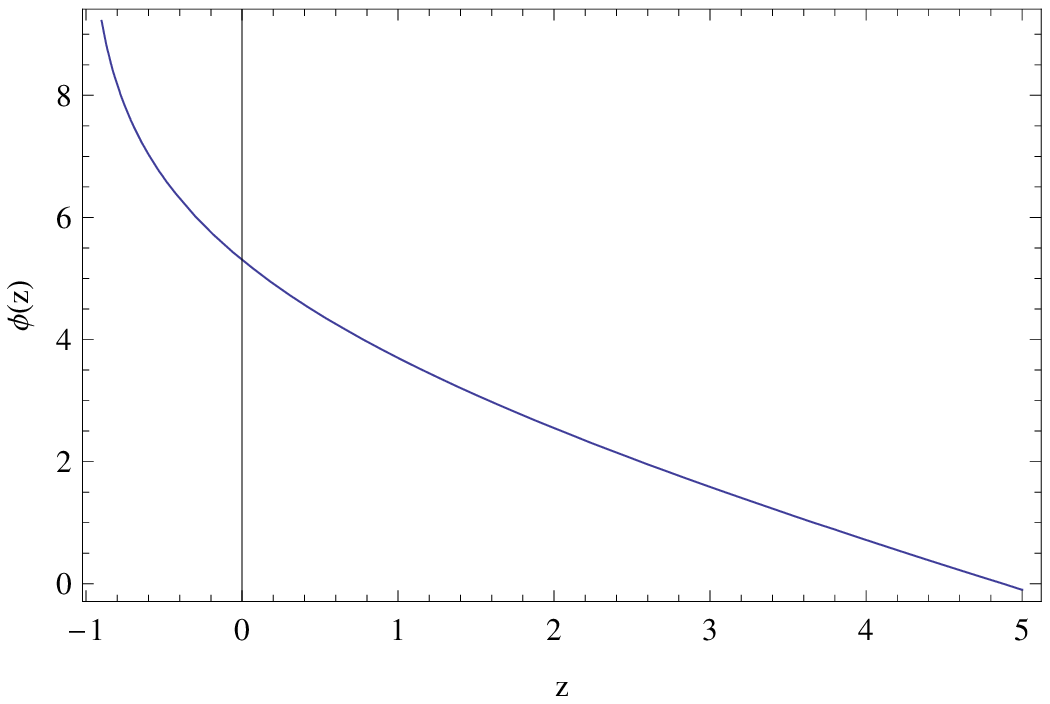}~~~~
\includegraphics[height=2.0in]{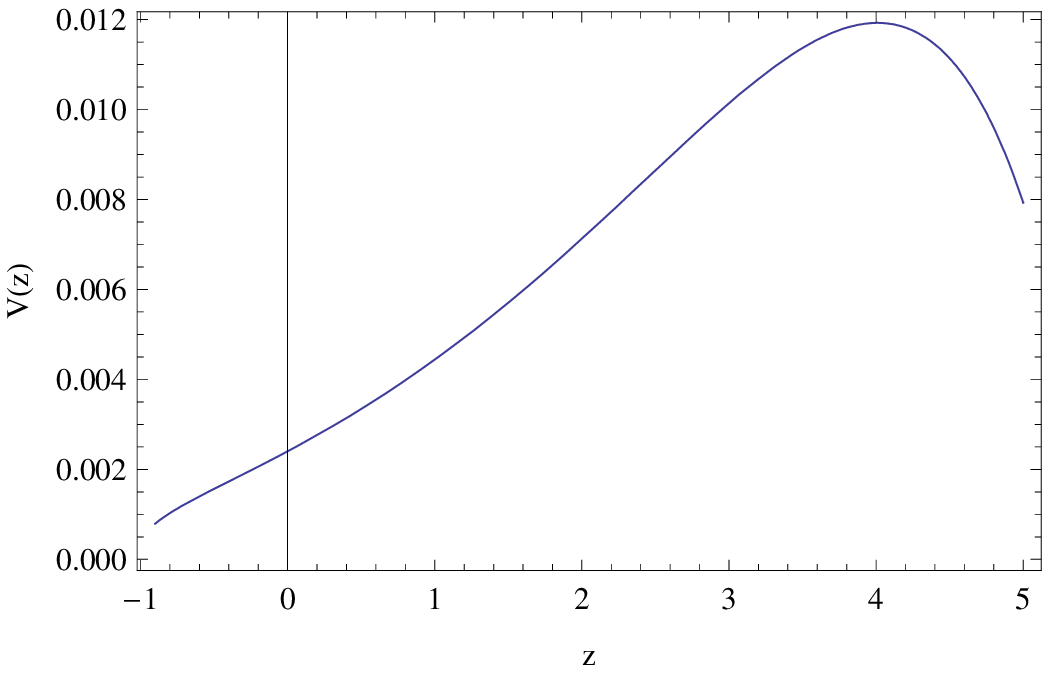}\\

~~~~~~~~~~~~~~~~~~~~~~~~~~~~~~~~~~Fig.5~~~~~~~~~~~~~~~~~~~~~~~~~~~~~~~~~~~~~~~~~~~~~~~~~~~~~~~~~~~~~~~~Fig.6\\
\vspace{2mm}

Fig. 1 shows the variations of $w_{0}$ and $w_{1}$ in the joint
analysis (SNIa+BAO+CMB+Hubble) for the {\bf linear}
parameterization (Model I). We plot the graphs for different
confidence levels 66\% (solid, blue), 90\% (dashed, red) and 99\%
(dashed, black) contours for ($w_{0},~w_{1}$) by fixing the other
parameters. Figs. 2, 3, 4, 5 and 6 show the variations of
$H,~q,~f,~\phi$ and $V$ with the variation in $z$. Here we have
chosen
$\beta=-0.5,D=5,w_{m}=-0.3,w_{0}=-0.737,w_{1}=0.173,\omega=0.2,
v_{0}=0.5,f_{0}=2,n=3,H_{0}=72,\Omega_{m0}=0.3,\Omega_{k0}=0.05$. \\

\vspace{4mm}

\end{figure}

\begin{figure}

\includegraphics[height=2.0in]{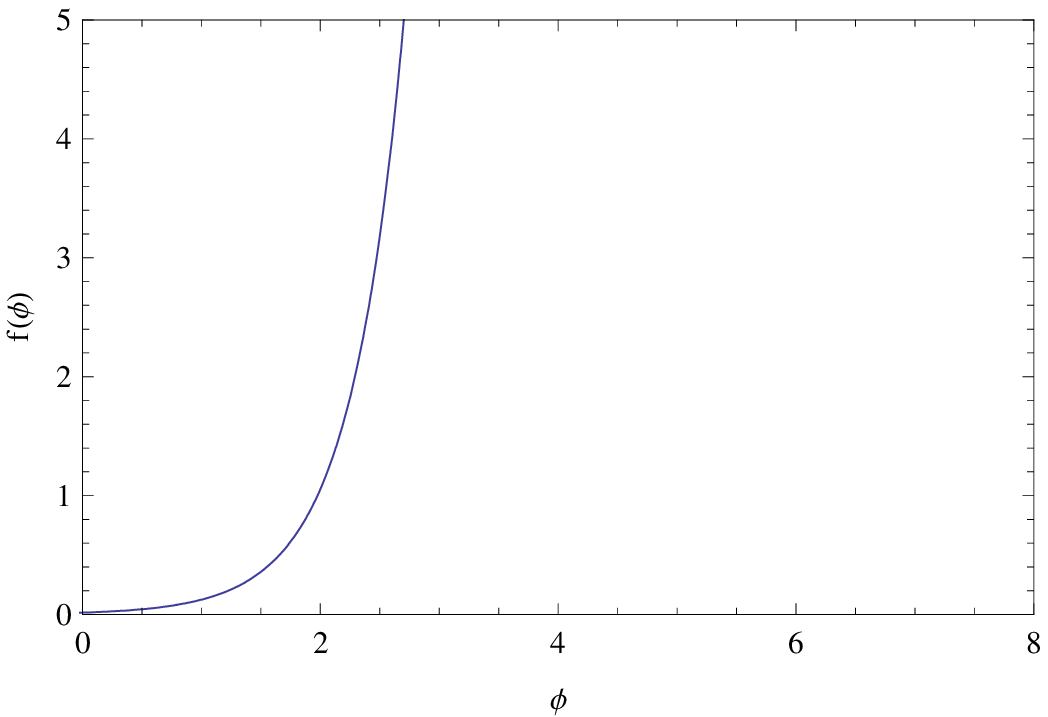}~~~
\includegraphics[height=2.0in]{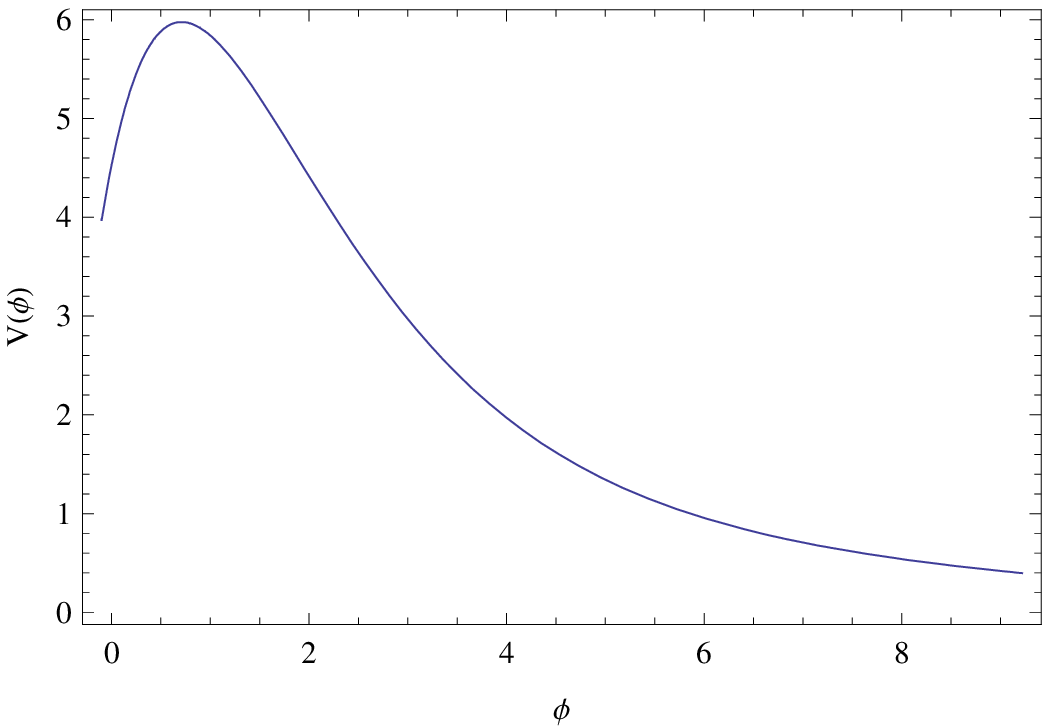}\\

~~~~~~~~~~~~~~~~~~~~~~~~~~~~~~~~~~Fig.7~~~~~~~~~~~~~~~~~~~~~~~~~~~~~~~~~~~~~~~~~~~~~~~~~~~~~~~~~~~~~~~~Fig.8\\

Figs. 7 and 8 show the variations of $f(\phi)$ and $V(\phi)$ with
the variation of $\phi$ for the {\bf linear} parameterization
(Model I). Here we have chosen
$\beta=-0.5,D=5,w_{m}=-0.3,w_{0}=-0.737,w_{1}=0.173,\omega=0.2,
v_{0}=0.5,f_{0}=2,n=3,H_{0}=72,\Omega_{m0}=0.3,\Omega_{k0}=0.05$. \\

\vspace{4mm}

\end{figure}

\subsection{\normalsize\bf{Model II : Chevallier-Polarski-Linder (CPL) Parameterization}}

In the Chevallier-Polarski-Linder (CPL) Parameterization model,
the equation of state parameter is given by \cite{Chev,Linder}
$w_{\phi}(z)=w_0+{w_1}\frac{z}{1+z}$. Here again $w_0$ and $w_1$
are constants in which $w_{0}$ represents the present value of
$w_{\phi}(z)$. With these, the expressions of energy density
becomes

\begin{equation}
\rho_{\phi}=\rho_{\phi
0}(1+z)^{(\beta+D-1)(1+w_0+{w_1})}e^{-\frac{(\beta+D-1){w_1}z}{1+z}}
\end{equation}

From equation (\ref{27}), the Hubble parameter can be written as

\begin{equation}\label{32}
H^{2}(z)=\frac{\xi
H_{0}^{2}\left[\Omega_{m0}(1+z)^{(\beta+D-1)(1+w_{m})}+\Omega_{\phi
0}~(1+z)^{(\beta+D-1)(1+w_0+{w_1})}e^{-\frac{(\beta+D-1){w_1}z}{1+z}}-\Omega_{k0}(1+z)^{2}\right]}{\left[\xi+\omega
v_{0}^{2}\beta^{2}-\omega
v_{0}^{2}\beta^{2}(1+z)^{-2\beta}\right]}
\end{equation}

\begin{figure}
\includegraphics[height=2.5in]{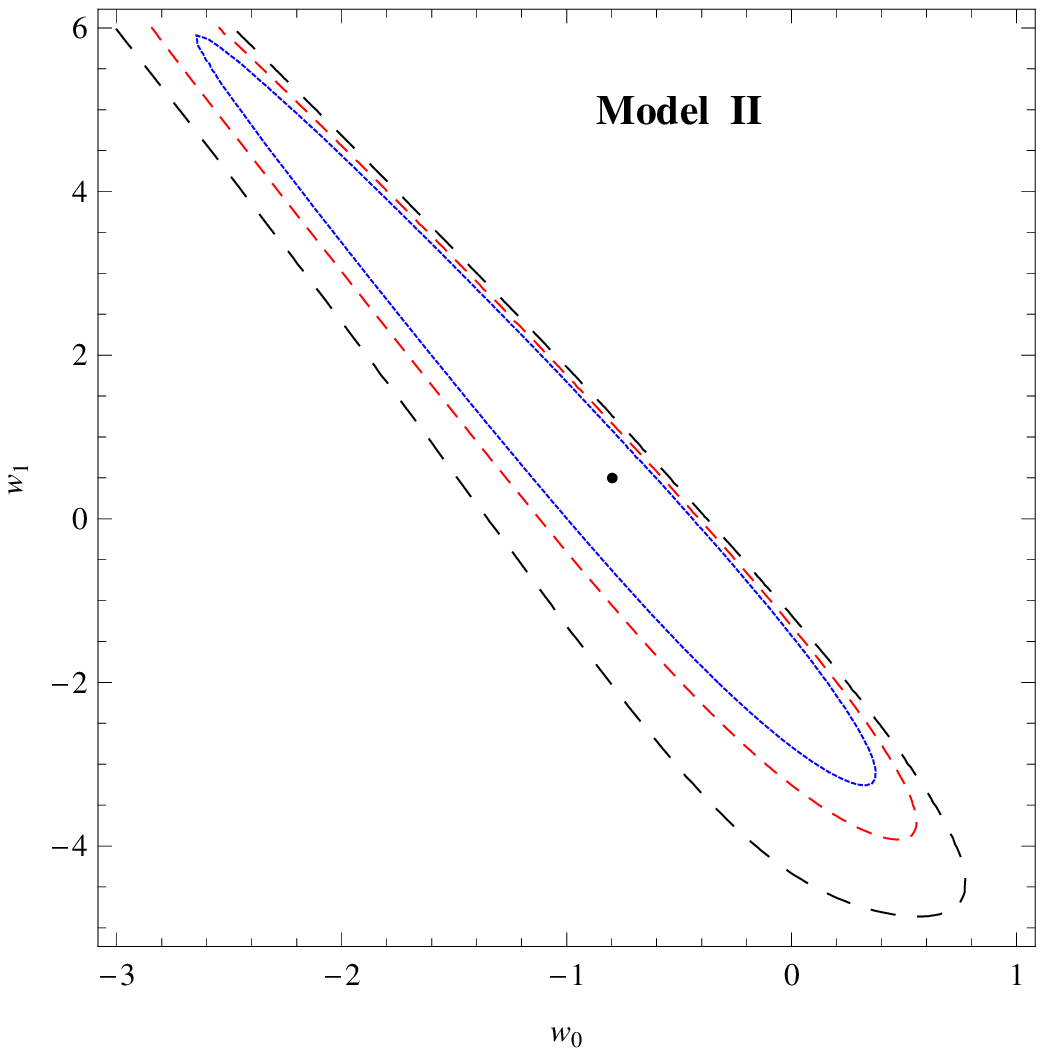}~~~~
\includegraphics[height=2.0in]{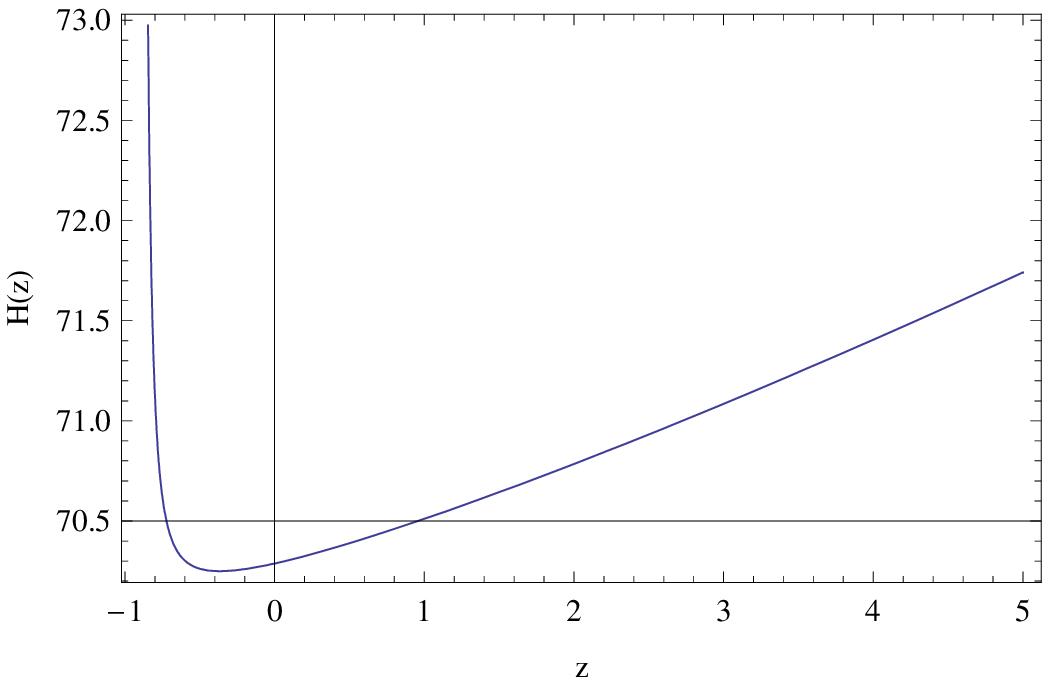}\\

~~~~~~~~~~~~~~~~~~~~~~~~~~~~~~~~~~Fig.9~~~~~~~~~~~~~~~~~~~~~~~~~~~~~~~~~~~~~~~~~~~~~~~~~~~~~~~~~~~~~~~~Fig.10\\
\vspace{2mm}

\includegraphics[height=2.0in]{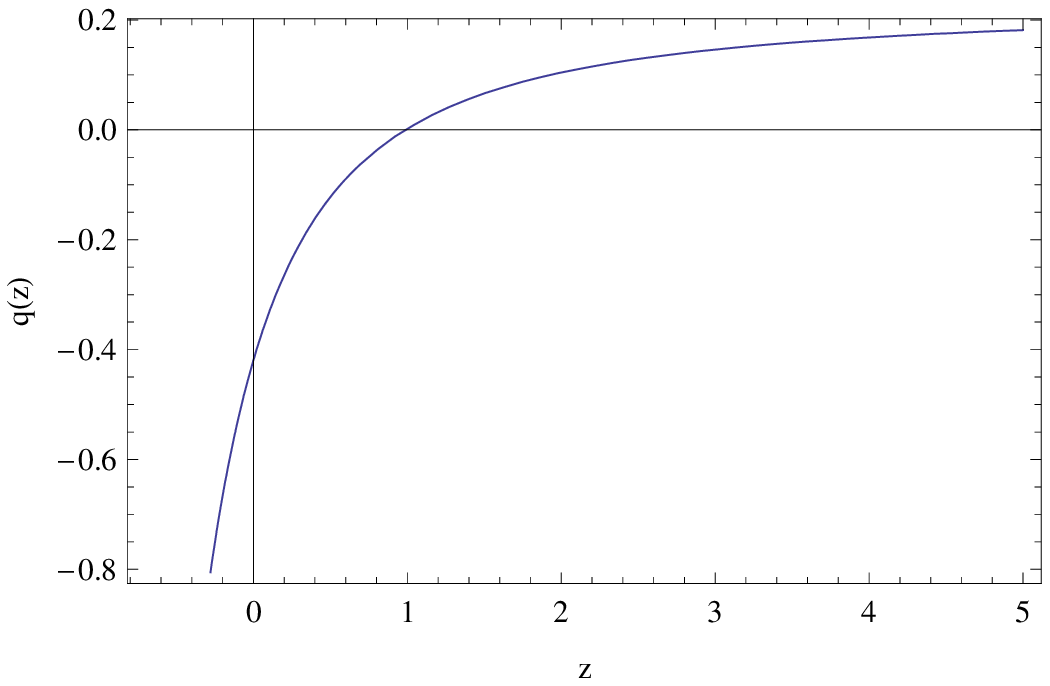}~~~~
\includegraphics[height=2.0in]{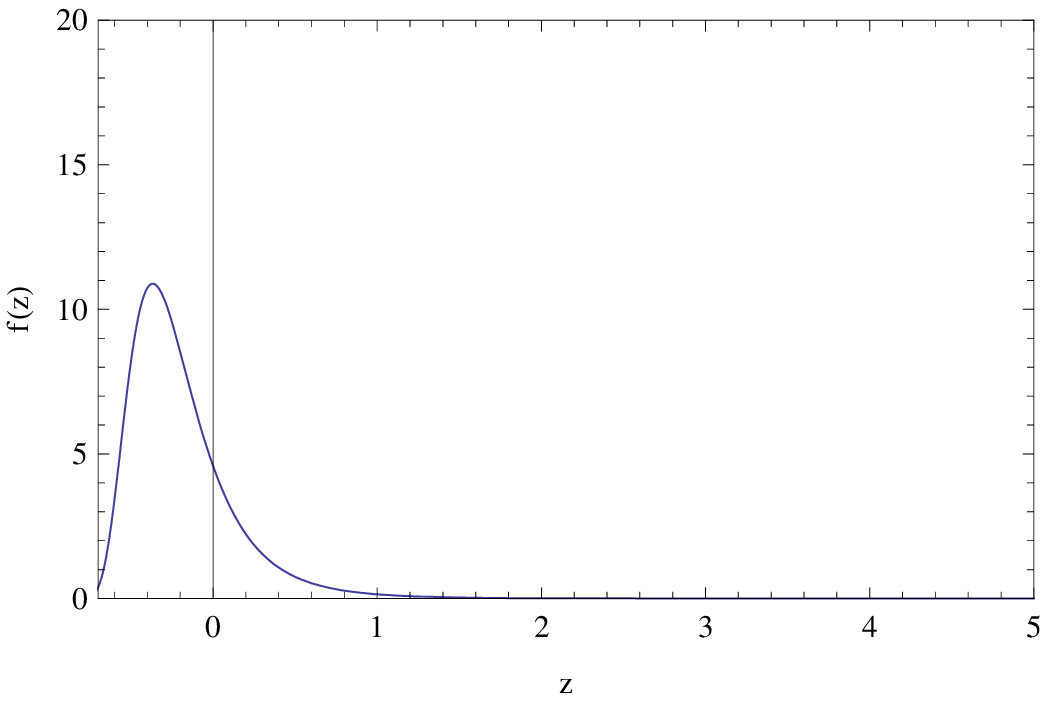}\\

~~~~~~~~~~~~~~~~~~~~~~~~~~~~~~~~~~Fig.11~~~~~~~~~~~~~~~~~~~~~~~~~~~~~~~~~~~~~~~~~~~~~~~~~~~~~~~~~~~~~~~~Fig.12\\
\vspace{2mm}

\includegraphics[height=2.0in]{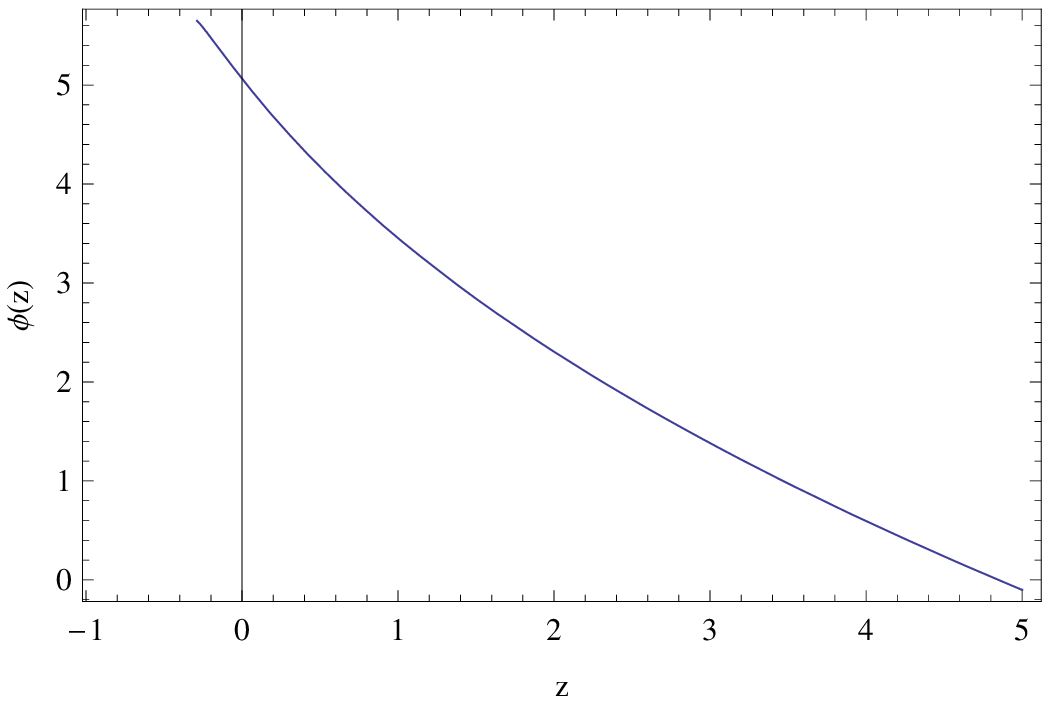}~~~~
\includegraphics[height=2.0in]{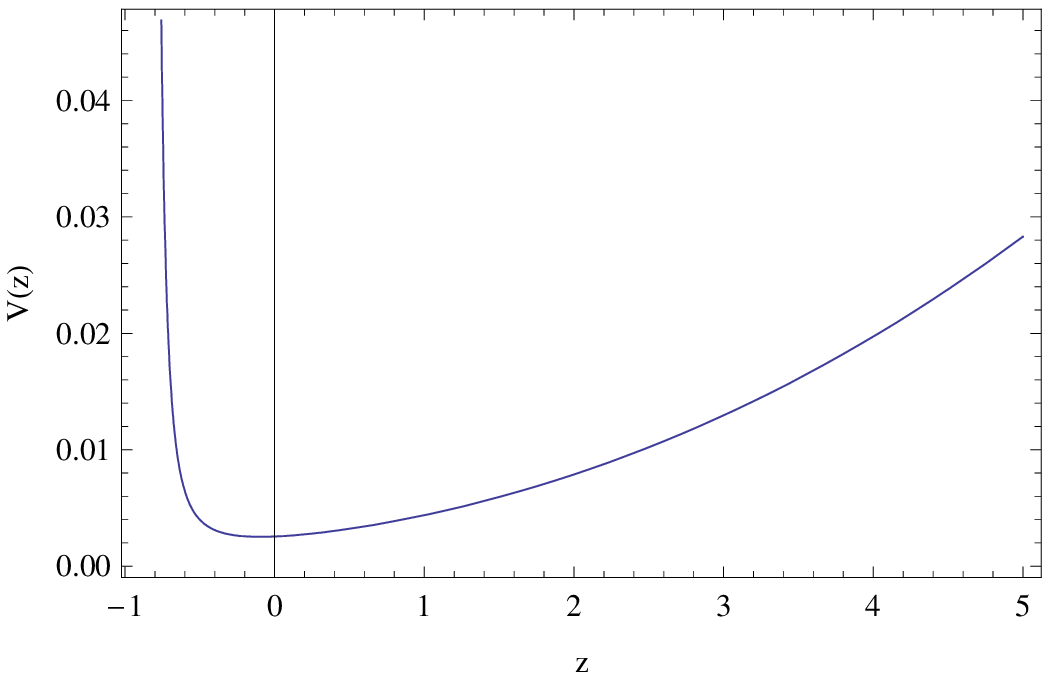}\\

~~~~~~~~~~~~~~~~~~~~~~~~~~~~~~~~~~Fig.13~~~~~~~~~~~~~~~~~~~~~~~~~~~~~~~~~~~~~~~~~~~~~~~~~~~~~~~~~~~~~~~~Fig.14\\
\vspace{2mm}

Fig. 9 shows the variations of $w_{0}$ and $w_{1}$ in the joint
analysis (SNIa+BAO+CMB+Hubble) for the {\bf CPL} parameterization
(Model II). We plot the graphs for different confidence levels
66\% (solid, blue), 90\% (dashed, red) and 99\% (dashed, black)
contours for ($w_{0},~w_{1}$) by fixing the other parameters.
Figs. 10, 11, 12, 13 and 14 show the variations of $H,~q,~f,~\phi$
and $V$ with the variation in $z$. Here we have chosen
$\beta=-0.5,D=5,w_{m}=-0.3,w_{0}=-0.797,w_{1}=0.499,\omega=0.2,
v_{0}=0.5,f_{0}=2,n=3,H_{0}=72,\Omega_{m0}=0.3,\Omega_{k0}=0.05$. \\

\vspace{4mm}

\end{figure}

\begin{figure}
\includegraphics[height=2.0in]{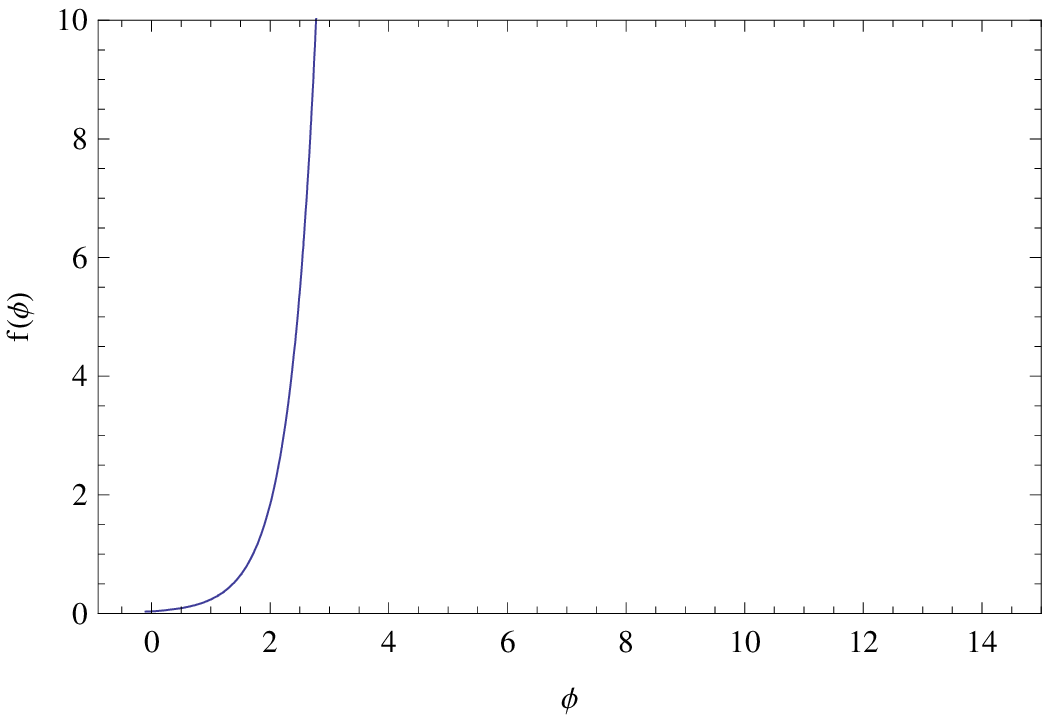}~~~~
\includegraphics[height=2.0in]{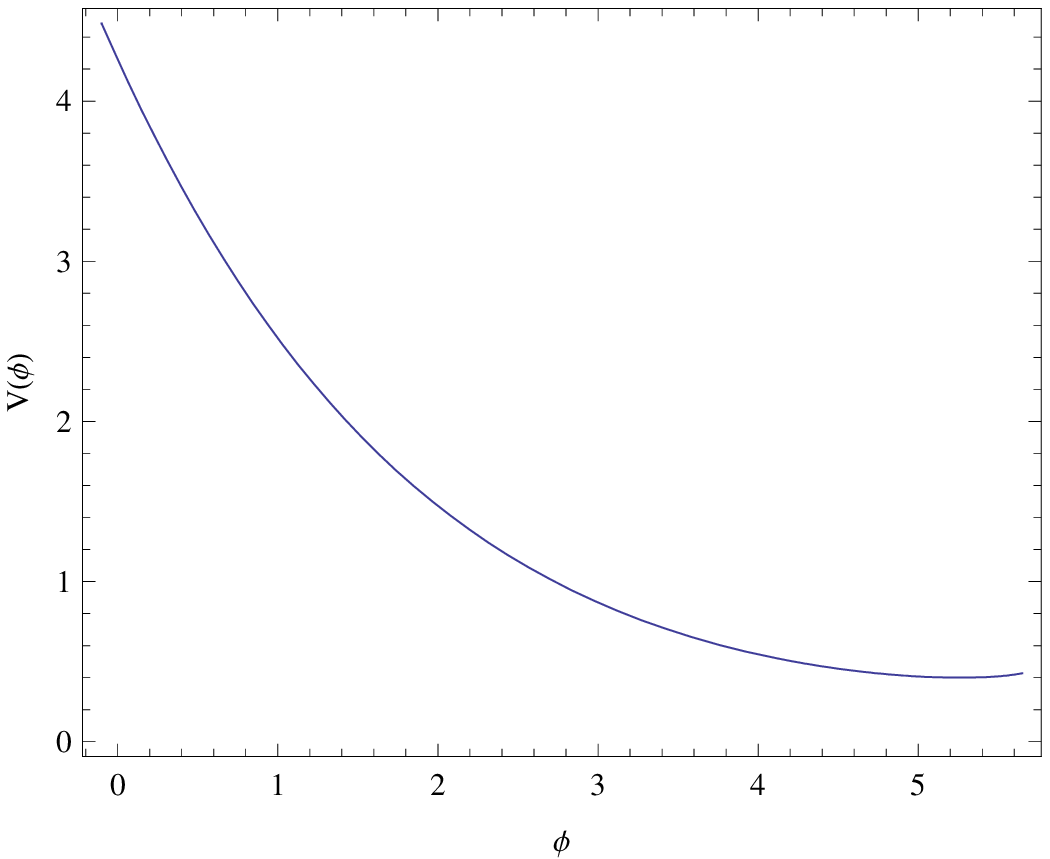}\\

~~~~~~~~~~~~~~~~~~~~~~~~~~~~~~~~~~Fig.15~~~~~~~~~~~~~~~~~~~~~~~~~~~~~~~~~~~~~~~~~~~~~~~~~~~~~~~~~~~~~~~~Fig.16\\

Figs. 15 and 16 show the variations of $f(\phi)$ and $V(\phi)$
with the variation of $\phi$ for the {\bf CPL} parameterization
(Model II). Here we have chosen
$\beta=-0.5,D=5,w_{m}=-0.3,w_{0}=-0.797,w_{1}=0.499,\omega=0.2,
v_{0}=0.5,f_{0}=2,n=3,H_{0}=72,\Omega_{m0}=0.3,\Omega_{k0}=0.05$. \\

\vspace{4mm}

\end{figure}

\subsection{\normalsize\bf{Model III : Jassal-Bagla-Padmanabhan (JBP) Parameterization}}

For the Jassal-Bagla-Padmanabhan (JBP) Parameterization model, the
equation of state parameter is \cite{Jassal}
$w_{\phi}(z)=w_0+{w_1}\frac{z}{(1+z)^2}$, where $w_0$ and $w_1$
are constants in which $w_{0}$ represents the present value of
$w_{\phi}(z)$. The following expression is subsequently obtained
as

\begin{equation}
\rho_{\phi}=\rho_{\phi
0}(1+z)^{(\beta+D-1)(1+w_0)}e^{\frac{(\beta+D-1){w_1}z^2}{2(1+z)^2}}
\end{equation}

From equation (\ref{27}), the Hubble parameter can be written as

\begin{equation}\label{34}
H^{2}(z)=\frac{\xi
H_{0}^{2}\left[\Omega_{m0}(1+z)^{(\beta+D-1)(1+w_{m})}+\Omega_{\phi
0}~(1+z)^{(\beta+D-1)(1+w_0)}e^{\frac{(\beta+D-1){w_1}z^2}{2(1+z)^2}}-\Omega_{k0}(1+z)^{2}\right]}{\left[\xi+\omega
v_{0}^{2}\beta^{2}-\omega
v_{0}^{2}\beta^{2}(1+z)^{-2\beta}\right]}
\end{equation}

\begin{figure}
\includegraphics[height=2.5in]{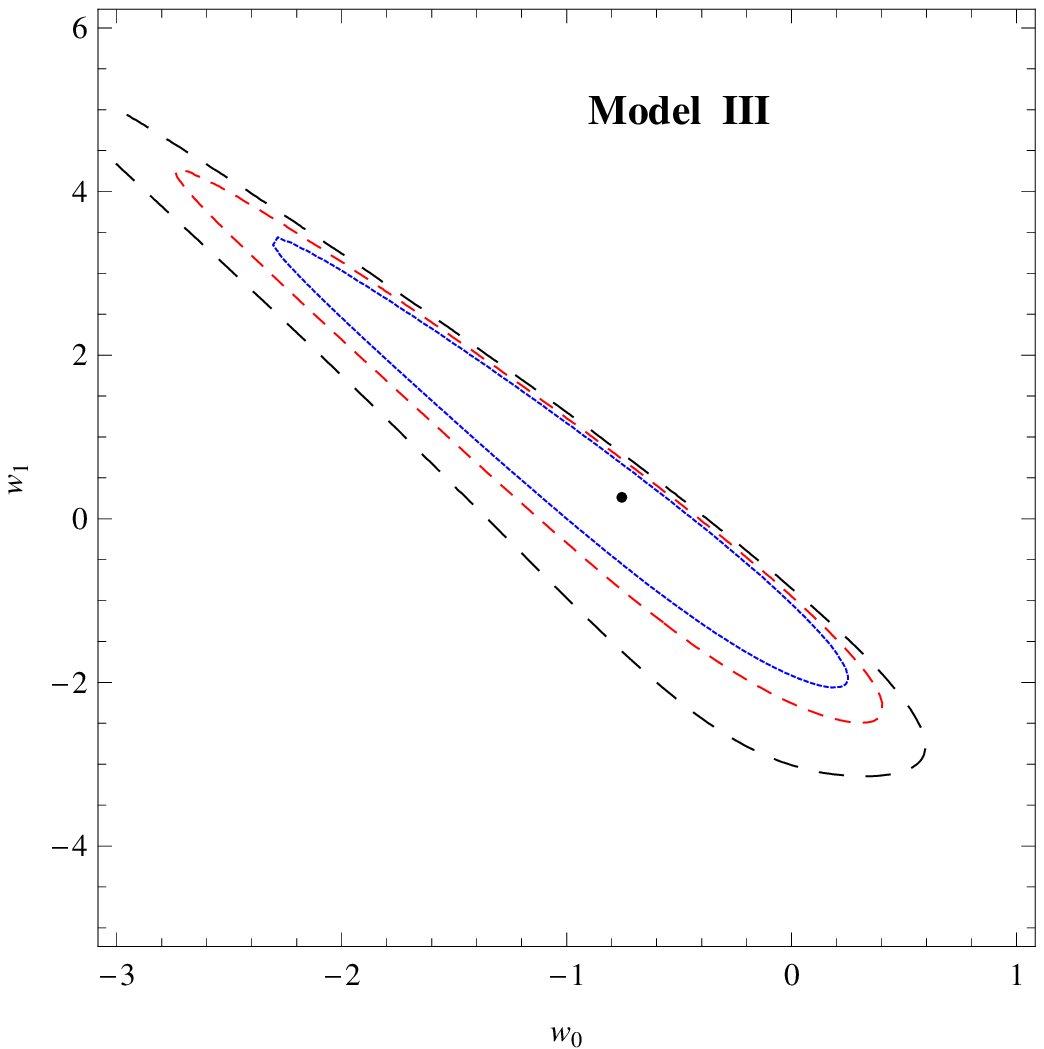}~~~~
\includegraphics[height=2.0in]{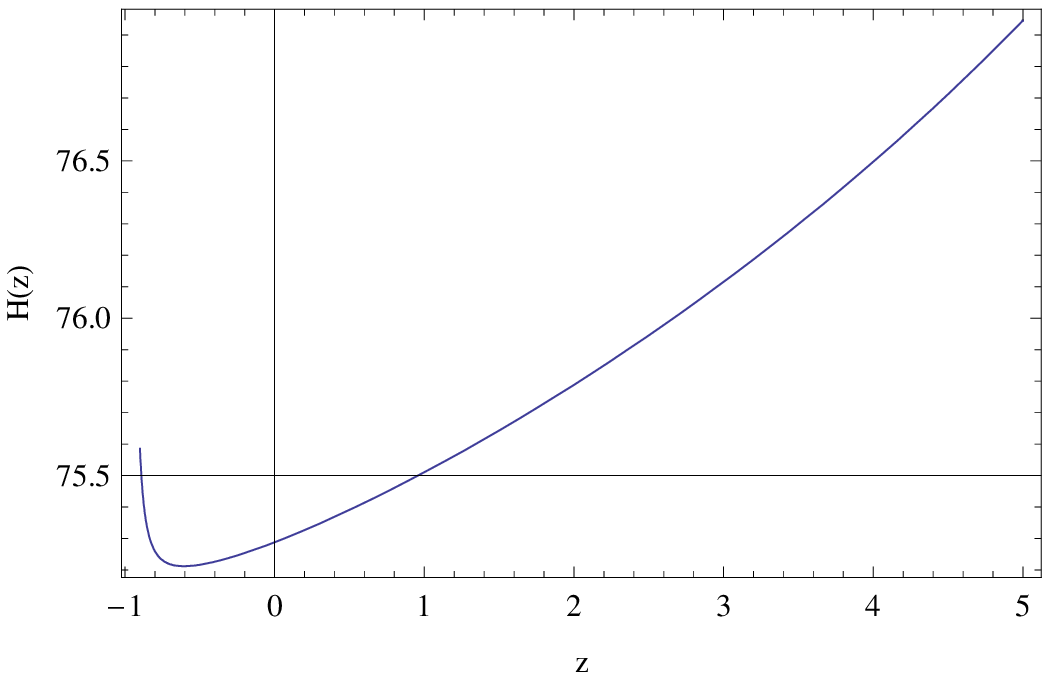}\\

~~~~~~~~~~~~~~~~~~~~~~~~~~~~~~~~~~Fig.17~~~~~~~~~~~~~~~~~~~~~~~~~~~~~~~~~~~~~~~~~~~~~~~~~~~~~~~~~~~~~~~~Fig.18\\
\vspace{2mm}

\includegraphics[height=2.0in]{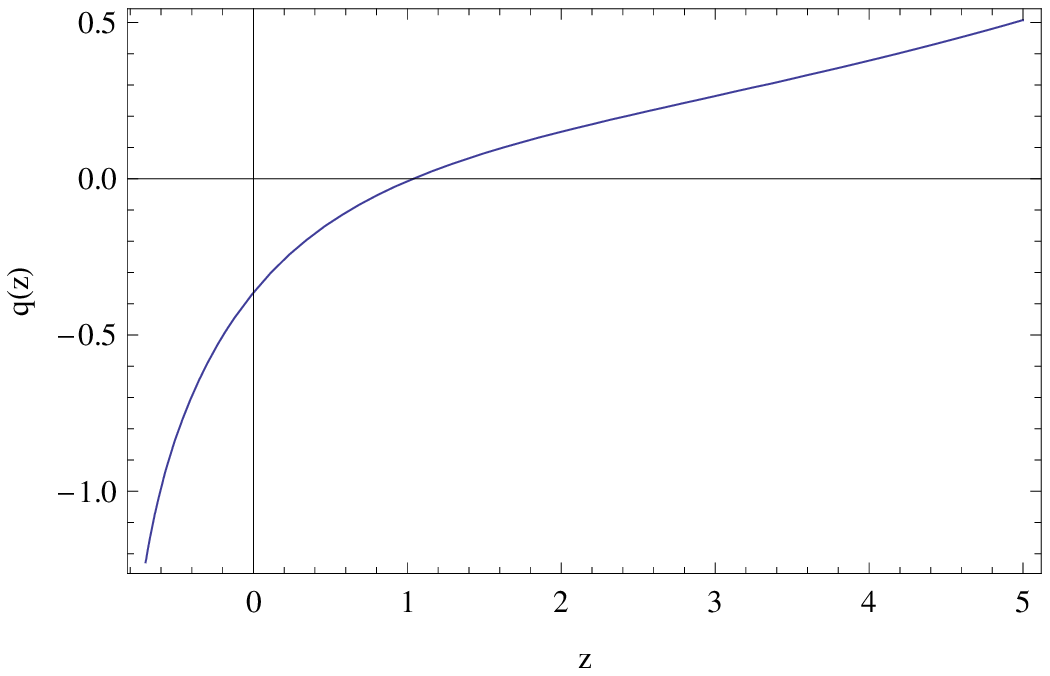}~~~~
\includegraphics[height=2.0in]{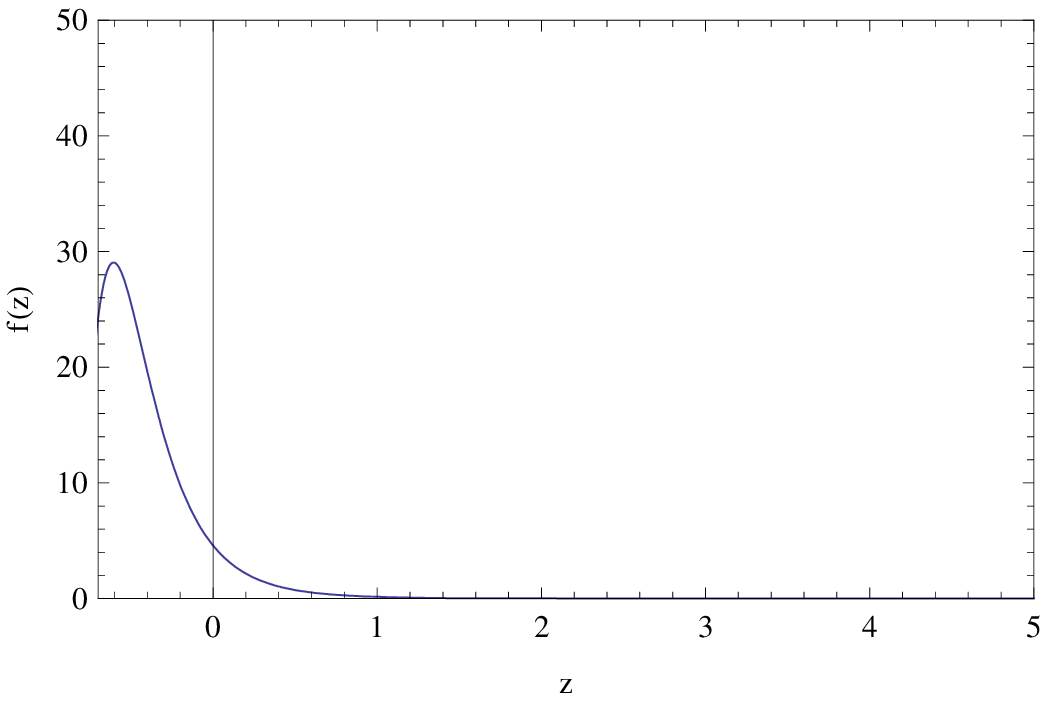}\\

~~~~~~~~~~~~~~~~~~~~~~~~~~~~~~~~~~Fig.19~~~~~~~~~~~~~~~~~~~~~~~~~~~~~~~~~~~~~~~~~~~~~~~~~~~~~~~~~~~~~~~~Fig.20\\
\vspace{2mm}

\includegraphics[height=2.0in]{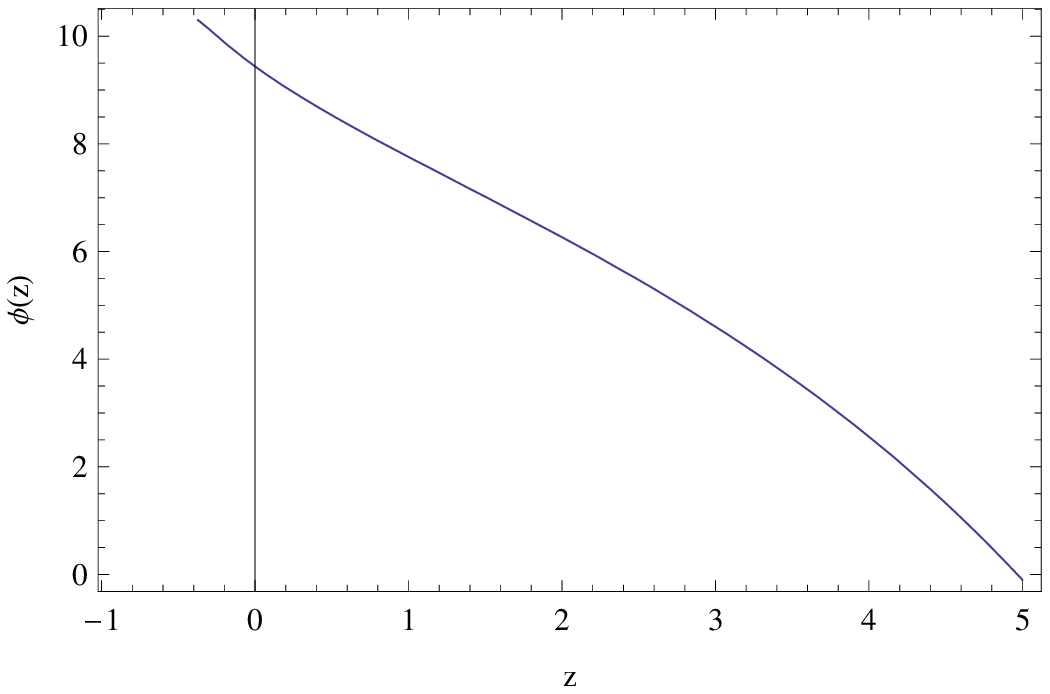}~~~~
\includegraphics[height=2.0in]{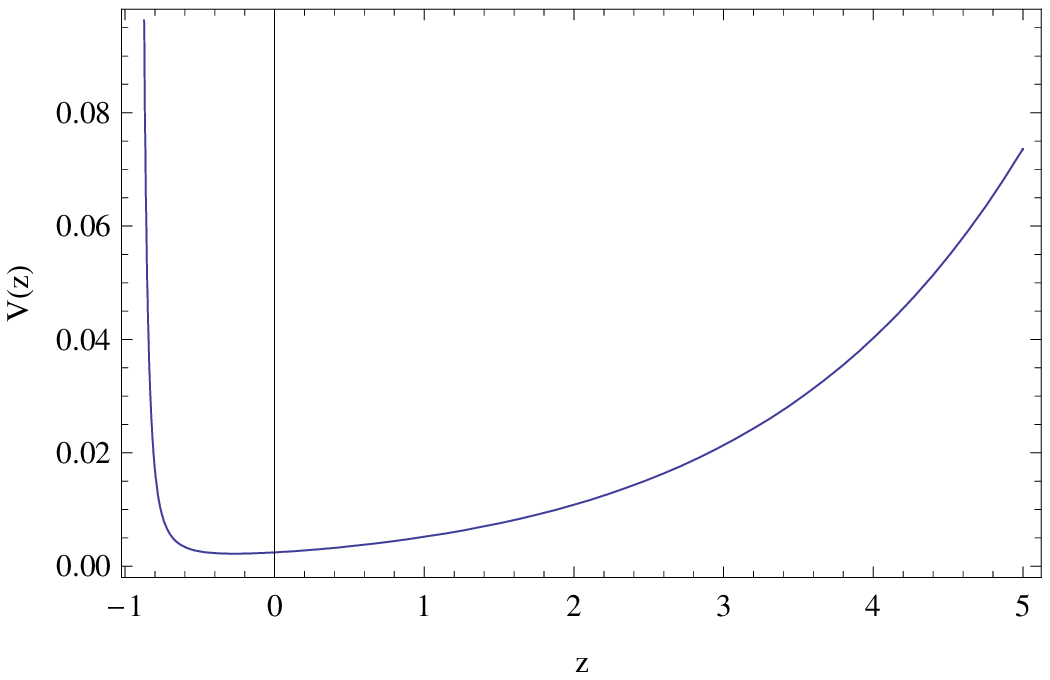}\\

~~~~~~~~~~~~~~~~~~~~~~~~~~~~~~~~~~Fig.21~~~~~~~~~~~~~~~~~~~~~~~~~~~~~~~~~~~~~~~~~~~~~~~~~~~~~~~~~~~~~~~~Fig.22\\
\vspace{2mm}

Fig. 17 shows the variations of $w_{0}$ and $w_{1}$ in the joint
analysis (SNIa+BAO+CMB+Hubble) for the {\bf JBP} parameterization
(Model III). We plot the graphs for different confidence levels
66\% (solid, blue), 90\% (dashed, red) and 99\% (dashed, black)
contours for ($w_{0},~w_{1}$) by fixing the other parameters.
Figs. 18, 19, 20, 21 and 22 show the variations of $H,~q,~f,~\phi$
and $V$ with the variation in $z$. Here we have chosen
$\beta=-0.5,D=5,w_{m}=-0.3,w_{0}=-0.754,w_{1}=0.261,\omega=0.2,
v_{0}=0.5,f_{0}=2,n=3,H_{0}=72,\Omega_{m0}=0.3,\Omega_{k0}=0.05$. \\

\vspace{4mm}

\end{figure}

\begin{figure}

\includegraphics[height=2.0in]{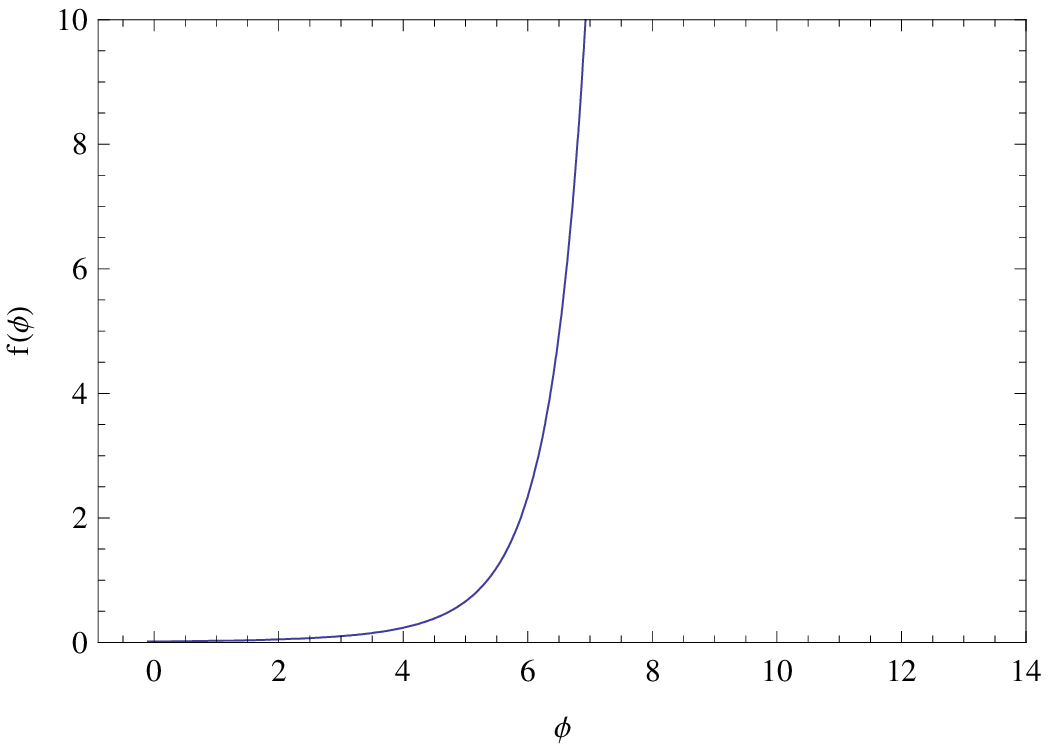}~~~~
\includegraphics[height=2.0in]{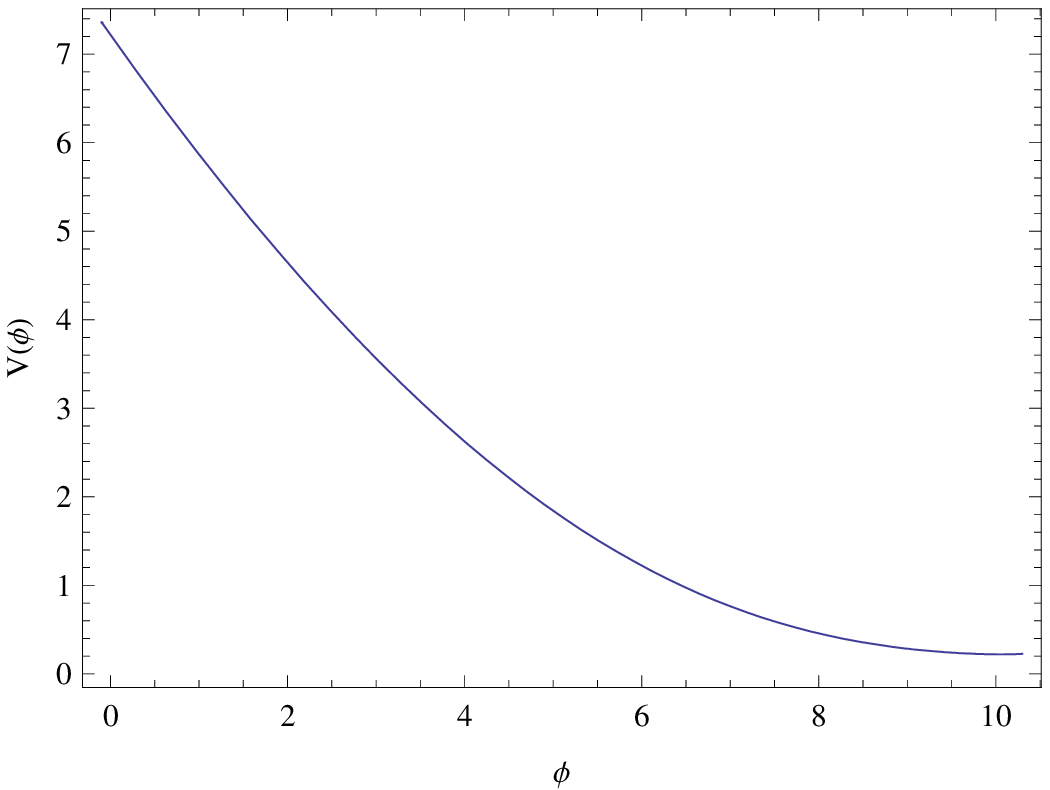}\\

~~~~~~~~~~~~~~~~~~~~~~~~~~~~~~~~~~Fig.23~~~~~~~~~~~~~~~~~~~~~~~~~~~~~~~~~~~~~~~~~~~~~~~~~~~~~~~~~~~~~~~~Fig.24\\

Figs. 23 and 24 show the variations of $f(\phi)$ and $V(\phi)$
with the variation of $\phi$ for the {\bf JBP} parameterization
(Model III). Here we have chosen
$\beta=-0.5,D=5,w_{m}=-0.3,w_{0}=-0.754,w_{1}=0.261,\omega=0.2,
v_{0}=0.5,f_{0}=2,n=3,H_{0}=72,\Omega_{m0}=0.3,\Omega_{k0}=0.05$. \\

\vspace{4mm}

\end{figure}

\subsection{\normalsize\bf{Model IV: Efstathiou
parametrization}}

Here, in Efstathiou parametrization model, the equation of state
parameter takes the form \cite{Ef,Sil}
$w_{\phi}(z)=w_0+w_1~log(1+z)$, where again $w_0$ and $w_1$ are
constants in which $w_{0}$ represents the present value of
$w_{\phi}(z)$. This gives rise to the following expression

\begin{equation}
\rho_{\phi}=\rho_{\phi
0}(1+z)^{(\beta+D-1)(1+w_0)}e^{\frac{(\beta+D-1)w_1}{2}[log(1+z)]^2}
\end{equation}

From equation (\ref{27}), the Hubble parameter can be written as

\begin{equation}\label{36}
H^{2}(z)=\frac{\xi
H_{0}^{2}\left[\Omega_{m0}(1+z)^{(\beta+D-1)(1+w_{m})}+\Omega_{\phi
0}~(1+z)^{(\beta+D-1)(1+w_0)}e^{\frac{(\beta+D-1)w_1}{2}[log(1+z)]^2}-\Omega_{k0}(1+z)^{2}\right]}{\left[\xi+\omega
v_{0}^{2}\beta^{2}-\omega
v_{0}^{2}\beta^{2}(1+z)^{-2\beta}\right]}
\end{equation}

\begin{figure}
\includegraphics[height=2.5in]{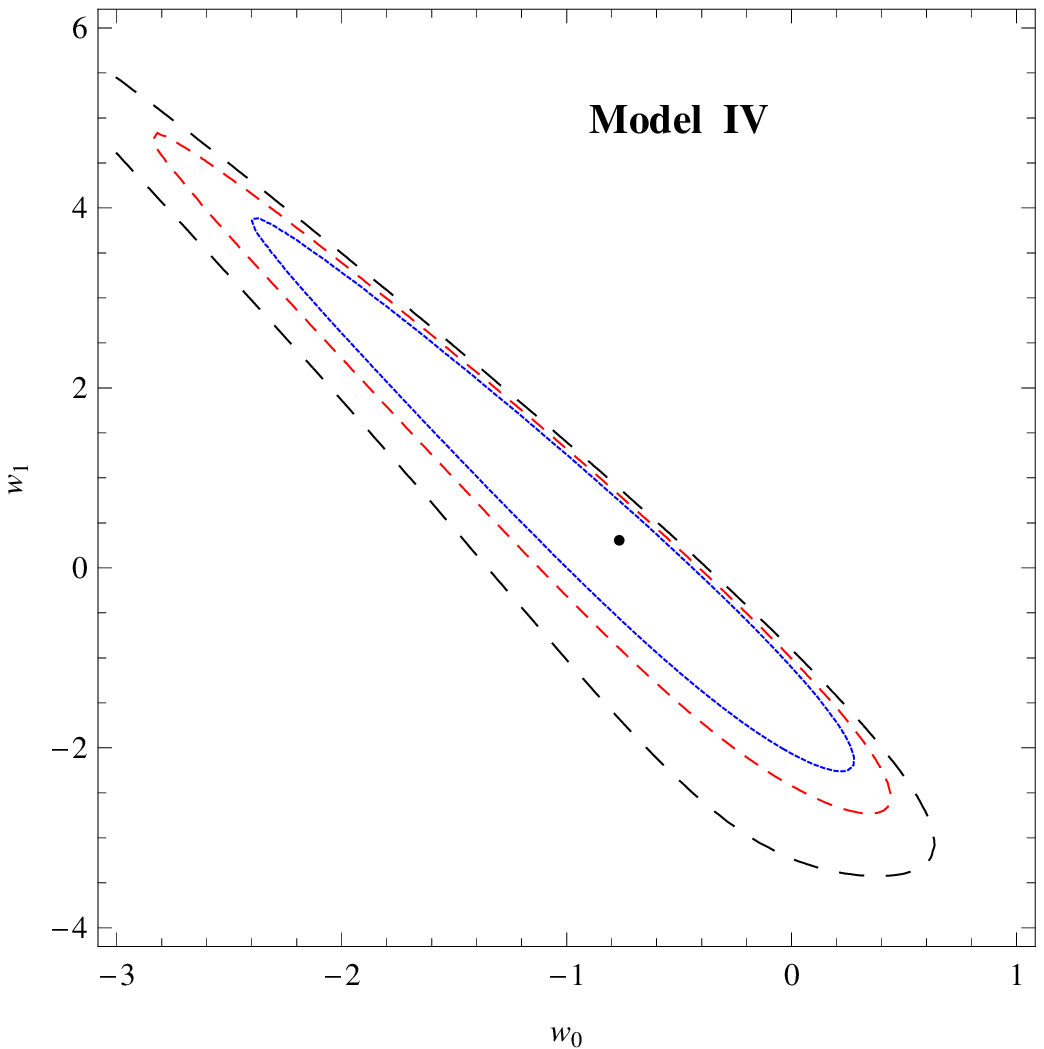}~~~~
\includegraphics[height=2.0in]{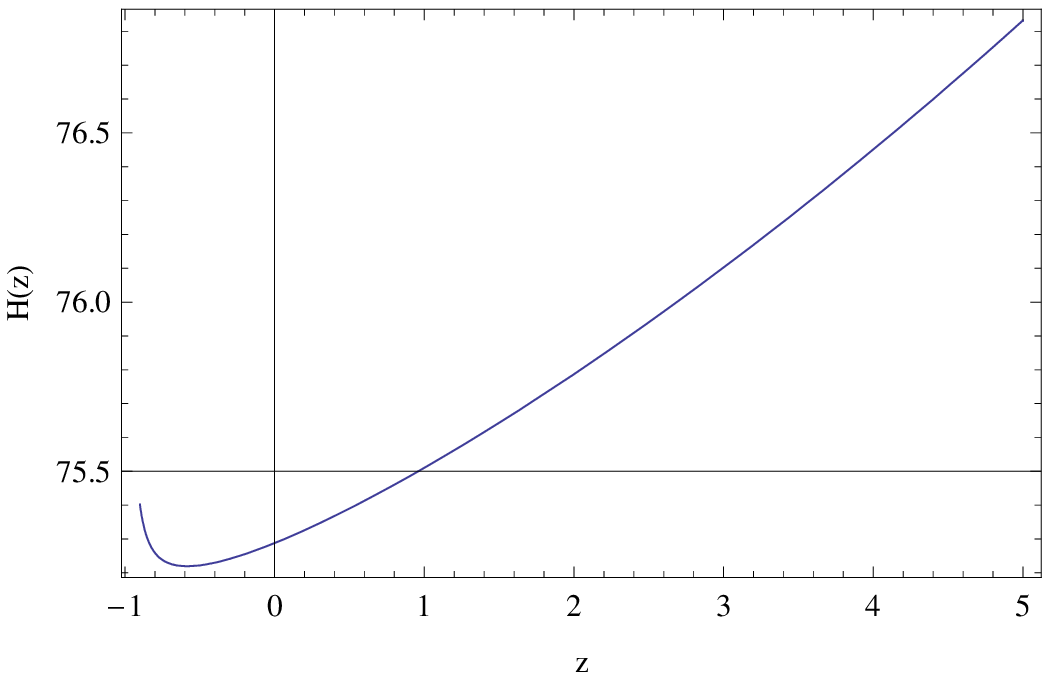}\\

~~~~~~~~~~~~~~~~~~~~~~~~~~~~~~~~~~Fig.25~~~~~~~~~~~~~~~~~~~~~~~~~~~~~~~~~~~~~~~~~~~~~~~~~~~~~~~~~~~~~~~~Fig.26\\
\vspace{2mm}

\includegraphics[height=2.0in]{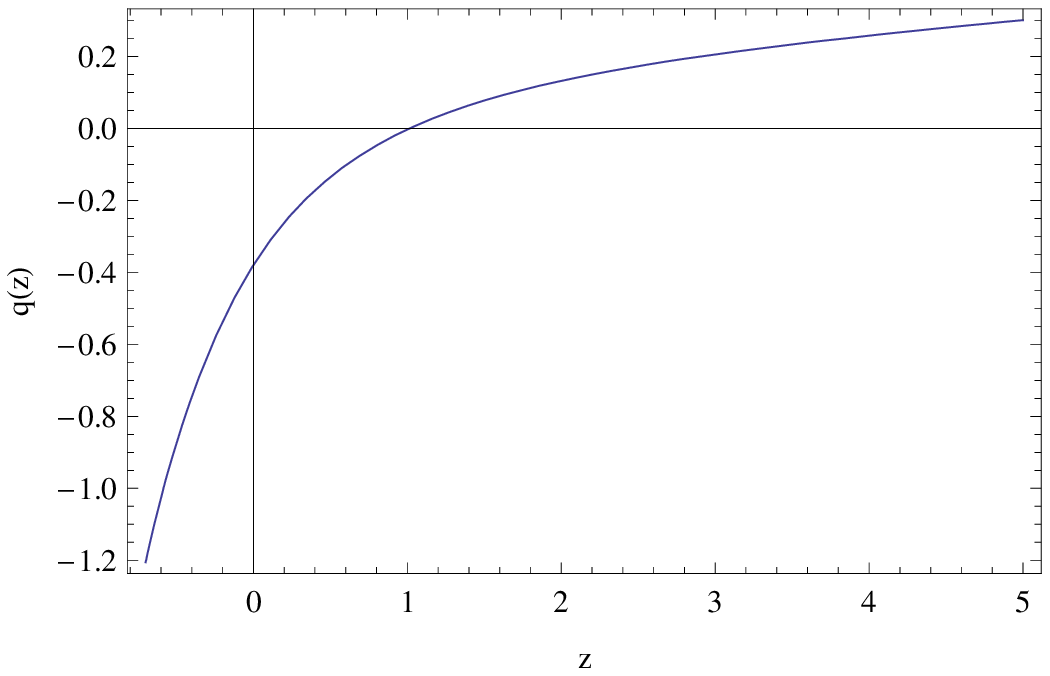}~~~~
\includegraphics[height=2.0in]{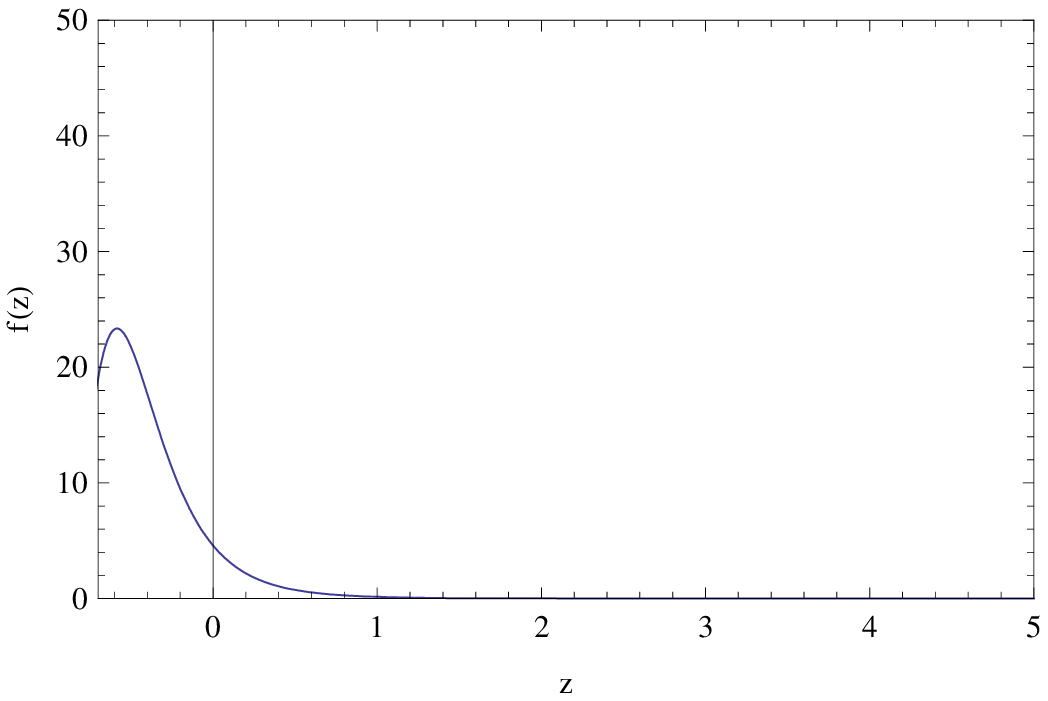}\\

~~~~~~~~~~~~~~~~~~~~~~~~~~~~~~~~~~Fig.27~~~~~~~~~~~~~~~~~~~~~~~~~~~~~~~~~~~~~~~~~~~~~~~~~~~~~~~~~~~~~~~~Fig.28\\
\vspace{2mm}

\includegraphics[height=2.0in]{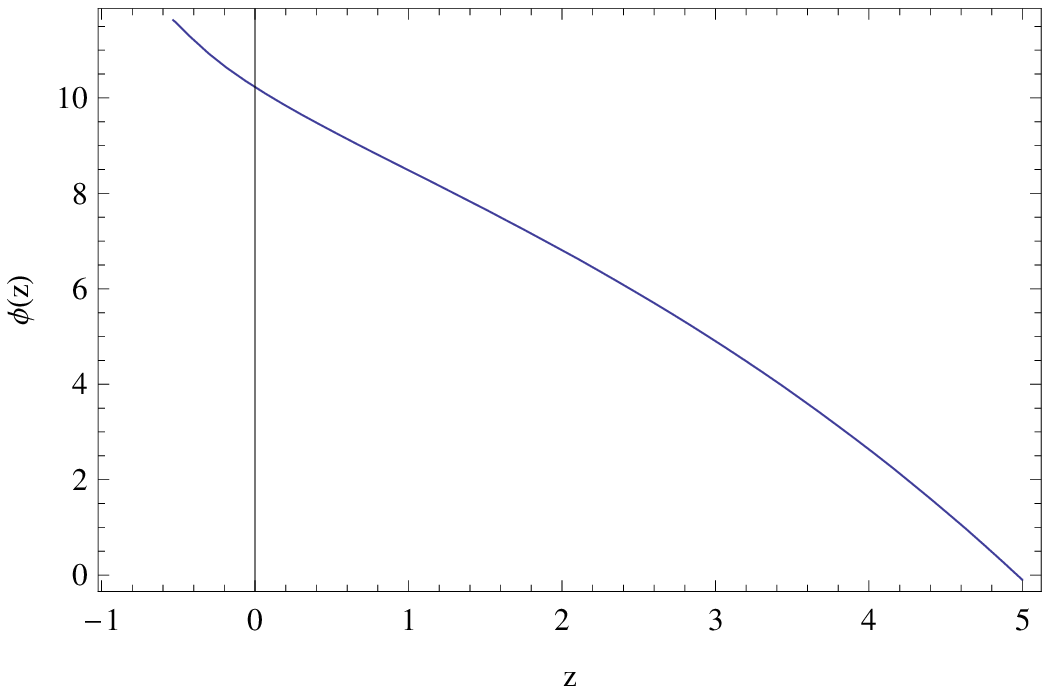}~~~~
\includegraphics[height=2.0in]{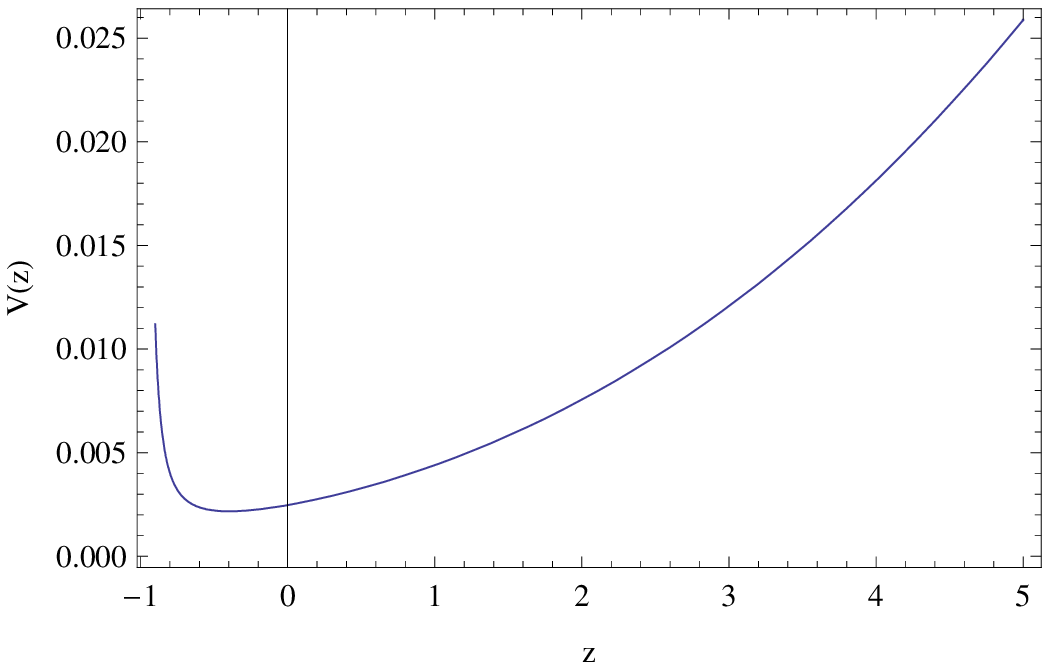}\\

~~~~~~~~~~~~~~~~~~~~~~~~~~~~~~~~~~Fig.29~~~~~~~~~~~~~~~~~~~~~~~~~~~~~~~~~~~~~~~~~~~~~~~~~~~~~~~~~~~~~~~~Fig.30\\
\vspace{2mm}

Fig. 25 shows the variations of $w_{0}$ and $w_{1}$ in the joint
analysis (SNIa+BAO+CMB+Hubble) for the {\bf Efstathiou}
parameterization (Model IV). We plot the graphs for different
confidence levels 66\% (solid, blue), 90\% (dashed, red) and 99\%
(dashed, black) contours for ($w_{0},~w_{1}$) by fixing the other
parameters. Figs. 26, 27, 28, 29 and 30 show the variations of
$H,~q,~f,~\phi$ and $V$ with the variation in $z$. Here we have
chosen
$\beta=-0.5,D=5,w_{m}=-0.3,w_{0}=-0.766,w_{1}=0.307,\omega=0.2,
v_{0}=0.5,f_{0}=2,n=3,H_{0}=72,\Omega_{m0}=0.3,\Omega_{k0}=0.05$. \\

\vspace{4mm}

\end{figure}

\begin{figure}

\includegraphics[height=2.0in]{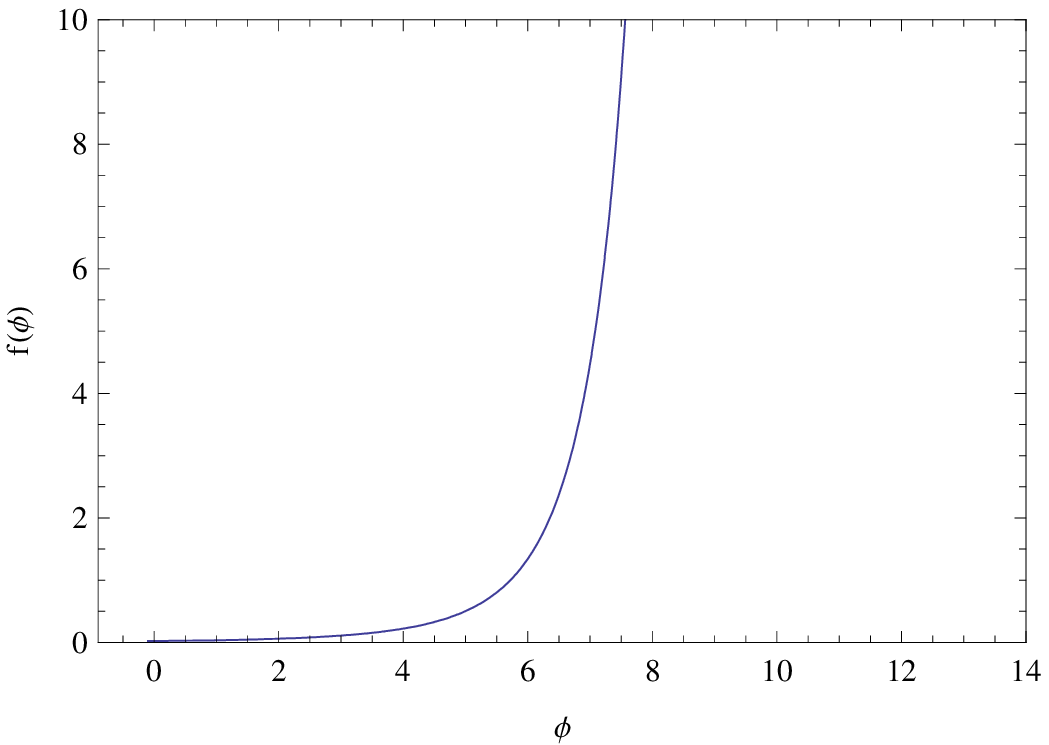}~~~~
\includegraphics[height=2.0in]{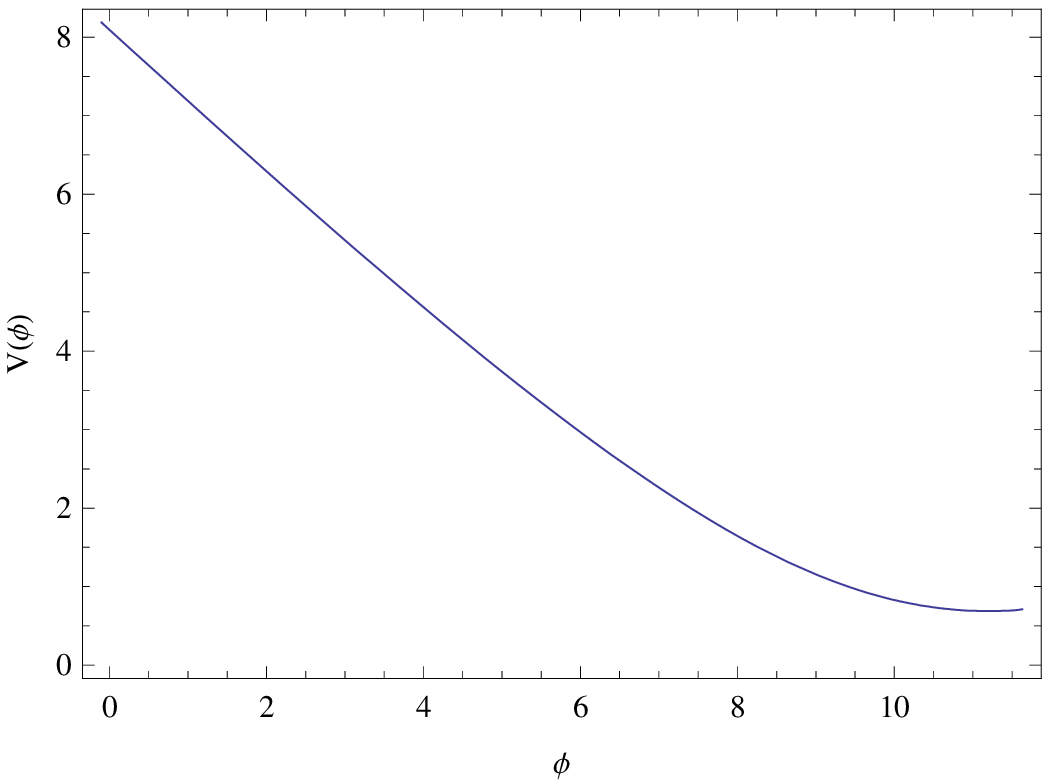}\\

~~~~~~~~~~~~~~~~~~~~~~~~~~~~~~~~~~Fig.31~~~~~~~~~~~~~~~~~~~~~~~~~~~~~~~~~~~~~~~~~~~~~~~~~~~~~~~~~~~~~~~~Fig.32\\

Figs. 31 and 32 show the variations of $f(\phi)$ and $V(\phi)$
with the variation of $\phi$ for the {\bf Efstathiou}
parameterization (Model IV). Here we have chosen
$\beta=-0.5,D=5,w_{m}=-0.3,w_{0}=-0.766,w_{1}=0.307,\omega=0.2,
v_{0}=0.5,f_{0}=2,n=3,H_{0}=72,\Omega_{m0}=0.3,\Omega_{k0}=0.05$. \\

\vspace{4mm}

\end{figure}

\section{Observational Data Analysis Technique}

In this section, we shall discuss the mechanism for fitting the
theoretical models with the recent observational data sets from
the type Ia supernova (SN Ia), the baryonic acoustic oscillations
(BAO) and the cosmic microwave background (CMB) data survey.

\subsection{Data Sets}

$\bullet$ {\bf SNIa data set:} Here we assume 580 data points of
type Ia supernovae with redshift ranging from 0.015 to 1.414.
Using this data set, the $\chi^{2}$ function is given by
\cite{Mamon1,Cai1}

\begin{equation}
\chi^{2}_{SN}=A_{SN}-\frac{B_{SN}^{2}}{C_{SN}}
\end{equation}
where
\begin{equation}
A_{SN}=\sum_{i=1}^{580}\frac{[\mu_{th}(z_{i})-\mu_{obs}(z_{i})]^{2}}{\sigma^{2}(z_{i})}~,
\end{equation}
\begin{equation}
B_{SN}=\sum_{i=1}^{580}\frac{[\mu_{th}(z_{i})-\mu_{obs}(z_{i})]}{\sigma^{2}(z_{i})}~,
\end{equation}
and
\begin{equation}
C_{SN}=\sum_{i=1}^{580}\frac{1}{\sigma^{2}(z_{i})}
\end{equation}
Here the distance modulus $\mu(z)$ for any SNIa at a redshift $z$
is given by
\begin{equation}
\mu(z)=5~
log_{10}\left[(1+z)\int_{0}^{z}\frac{dz'}{E(z')}\right]+\mu_{0}
\end{equation}
where $\mu_{0}$ is a nuisance parameter which should be
marginalized. Also $\mu_{th}$ represents the theoretical distance
modulus, while $\mu_{obs}$ is the theoretical distance modulus and
$\sigma$ is the standard error associated with the data point and $E(z)=H(z)/H_{0}$ is the
normalized Hubble parameter.\\

$\bullet$ {\bf SNIa + BAO data set:} Eisenstein et al
\cite{Eisenstein} proposed the Baryon Acoustic Oscillation (BAO)
peak parameter. The BAO signal has been directly detected at a
scale  $\sim$ 100 MPc by SDSS survey. We shall investigate the
parameters of the prescribed models using the BAO peak joint
analysis for the redshift which has the range $0 < z < z_{1}$
where $z_1 = 0.35$. For the SDSS data sample, $z_{1}$ is called
the typical redshift which has been used in the early times
\cite{Doran}. The BAO peak parameter can be defined in the
following form:

\begin{eqnarray}
{\cal{A}}=\frac{\sqrt{\Omega_m}}{E(z_1)^{1/3}}\left(\frac{\int_{0}^{z_1}\frac{dz}{E(z)}}{z_1}\right)^{2/3}
\end{eqnarray}
where
\begin{eqnarray}
\Omega_m=\Omega_{m0} (1+z_1)^3 E(z_1)^{-2}
\end{eqnarray}

Using SDSS data set \cite{Eisenstein}, the value of ${\cal{A}}$ is
$0.469 \pm 0.017$ for flat model of the FRW universe. Now for BAO
analysis, the $\chi^2$ function can be written as in the following
form

\begin{eqnarray}
\chi^2_{BAO}=\frac{({\cal{A}}-0.469)^2}{(0.017)^2}
\end{eqnarray}

$\bullet$ {\bf SNIa + BAO + CMB data set:} For Cosmic Microwave
Background (CMB), the shift parameter of CMB power spectrum peak
is given by \cite{Elgaroy,Efstathiou}

\begin{eqnarray}
{\cal {R}}=\sqrt{\Omega_m} \int_{0}^{z_2} \frac{dz'}{E(z')}
\end{eqnarray}

where $z_2$ is the value of the redshift at the surface of last
scattering. The WMAP data gives the value ${\cal {R}}=1.726 \pm
0.018 $ at the redshift $z=1091.3$. For CMB measurement, the
$\chi^2$ function can be defined as

\begin{eqnarray}
\chi^2_{CMB}=\frac{({\cal {R}}-1.726)^2}{(0.018)^2}
\end{eqnarray}

$\bullet$ {\bf Hubble data set:} Here we use the observed Hubble
data set by Stern et al \cite{Stern} at different redshifts at 12
data points \cite{Sim,Gaz}. The $\chi^2$ function is given by

\begin{eqnarray}
\chi^2_{H}=\sum_{i=1}^{12}
\frac{\left(H(z_{i})-H_{obs}(z_{i})\right)^2}{\sigma^{2}(z_{i})}
\end{eqnarray}
where the redshift of these data falls in the region $0<z<1.75$.\\

The total joint data analysis (SNIa+BAO+CMB+Hubble) for the
$\chi^2$ function is defined by
\begin{eqnarray}
\chi^2_{Tot}=\chi^2_{SN}+\chi^2_{BAO}+\chi^2_{CMB}+\chi^{2}_{H}
\end{eqnarray}

The best fit values of the model parameters can be determined by
minimizing the corresponding chi-square value which is equivalent
to the maximum likelihood analysis.


\subsection{Data Fittings and Numerical Results}

$\bullet$ {\bf Model I (Linear):} For this model, using
SNIa+BAO+CMB+Hubble joint analysis, we found the minimum value of
$\chi^{2}_{Tot}=7.104$ and the best fit values of the parameters
$w_{0}=-0.738$ and $w_{1}=0.174$ where we have fixed the other
parameters $\beta=-0.5,D=5,w_{m}=-0.3,\omega=0.2,
v_{0}=0.5,f_{0}=2,n=3,\Omega_{m0}=0.3,\Omega_{k0}=0.05$ and
$H_{0}=72$ km s$^{-1}$ MPc$^{-1}$. We have plotted the contours of
($w_{0},~w_{1}$) in figure 1 for different confidence levels 66\%
(solid, blue), 90\% (dashed, red) and 99\% (dashed, black). Now
taking best fit values of the parameters $w_{0}$ and $w_{1}$, we
have drawn the Hubble parameter $H(z)$ vs redshift $z$ in figure 2
and the deceleration parameter $q(z)$ vs $z$ in figure 3. We have
seen that the Hubble parameter and deceleration parameter decrease
over the expansion of time. The deceleration parameter $q(z)$ has
a sign flip from positive to negative, so our model generates
deceleration phase to acceleration phase of the Universe. Also at
the present value of $z=0$, the deceleration parameter $q$ is
negative, so at present our Universe is undergoing the
acceleration phase. The function $f(z)$, non-canonical scalar
field $\phi(z)$ and its potential $V(z)$ have been drawn in
figures 4 - 6 respectively. We have seen that $f(z)$ and $\phi(z)$
increase as $z$ decreases but $V(z)$ first increases and then
decreases as $z$ decreases. Also $f(\phi)$ and $V(\phi)$ in terms
of scalar field $\phi$ have been drawn in figures 7 and 8
respectively. Here $f(\phi)$ increases but $V(\phi)$ first
increases and then decreases as $\phi$
increases.\\

$\bullet$ {\bf Model II (CPL):} Due to joint analysis of
SNIa+BAO+CMB+Hubble, we have found the minimum value of
$\chi^{2}_{Tot}=7.042$ and the best fit values of the parameters
$w_{0}=-0.796$ and $w_{1}=0.498$ where we have fixed the other
parameters $\beta=-0.5,D=5,w_{m}=-0.3,\omega=0.2,
v_{0}=0.5,f_{0}=2,n=3,\Omega_{m0}=0.3,\Omega_{k0}=0.05$ and
$H_{0}=72$ km s$^{-1}$ MPc$^{-1}$. We have plotted the contours of
the parameters ($w_{0},~w_{1}$) in figure 9 for different
confidence levels 66\% (solid, blue), 90\% (dashed, red) and 99\%
(dashed, black). Now taking best fit values of the parameters
$w_{0}$ and $w_{1}$, we have drawn the Hubble parameter $H(z)$ vs
redshift $z$ in figure 10 and the deceleration parameter $q(z)$ vs
$z$ in figure 11. We have seen that the Hubble parameter first
decreases and then increases and deceleration parameter decreases
over the evolution of the Universe. The deceleration parameter
$q(z)$ has a sign flip from positive to negative levels. The
function $f(z)$, non-canonical scalar field $\phi(z)$ and its
potential $V(z)$ have been drawn in figures 12 - 14 respectively.
We have seen that $f(z)$ first increases and thereafter decreases
and $\phi(z)$ increase as $z$ decreases but $V(z)$ first decreases
and then increases as $z$ decreases. Also $f(\phi)$ and $V(\phi)$
in terms of scalar field $\phi$ have been drawn in figures 15 and
16 respectively. Here $f(\phi)$ sharply increases but $V(\phi)$
decreases as $\phi$ increases.\\

$\bullet$ {\bf Model III (JBP):} For this model, by investigating
the SNIa+BAO+CMB+Hubble joint analysis, we have found the minimum
value of $\chi^{2}_{Tot}=7.088$ and the best fit values of the
parameters $w_{0}=-0.755$ and $w_{1}=0.262$ where we have fixed
the other parameters $\beta=-0.5,D=5,w_{m}=-0.3,\omega=0.2,
v_{0}=0.5,f_{0}=2,n=3,\Omega_{m0}=0.3,\Omega_{k0}=0.05$ and
$H_{0}=72$ km s$^{-1}$ MPc$^{-1}$. We have plotted the contours of
($w_{0},~w_{1}$) in figure 17 for different confidence levels 66\%
(solid, blue), 90\% (dashed, red) and 99\% (dashed, black). Now
taking best fit values of the parameters $w_{0}$ and $w_{1}$, we
have drawn the Hubble parameter $H(z)$ vs redshift $z$ in figure
18 and the deceleration parameter $q(z)$ vs $z$ in figure 19. We
have seen that the Hubble parameter first decreases and then
increases and deceleration parameter decreases as $z$ decreases.
The deceleration parameter $q(z)$ has a sign flip from positive to
negative and passes phantom barrier. The function $f(z)$,
non-canonical scalar field $\phi(z)$ and its potential $V(z)$ have
been drawn in figures 20 - 22 respectively. We have seen that
$f(z)$ and $\phi(z)$ increase as $z$ decreases but $V(z)$ first
decreases and then sharply increases as $z$ decreases. Also
$f(\phi)$ and $V(\phi)$ in terms of scalar field $\phi$ have been
drawn in figures 23 and 24 respectively. Here $f(\phi)$ sharply
increases but $V(\phi)$ decreases as $\phi$
increases.\\

$\bullet$ {\bf Model IV (Efstathiou):} Using SNIa+BAO+CMB+Hubble
joint analysis, we have found the minimum value of
$\chi^{2}_{Tot}=7.072$ and the best fit values of the parameters
$w_{0}=-0.765$ and $w_{1}=0.308$ where we have fixed the other
parameters $\beta=-0.5,D=5,w_{m}=-0.3,\omega=0.2,
v_{0}=0.5,f_{0}=2,n=3,\Omega_{m0}=0.3,\Omega_{k0}=0.05$ and
$H_{0}=72$ km s$^{-1}$ MPc$^{-1}$. We have plotted the contours of
($w_{0},~w_{1}$) in figure 25 for different confidence levels 66\%
(solid, blue), 90\% (dashed, red) and 99\% (dashed, black). Now
taking best fit values of the parameters $w_{0}$ and $w_{1}$, we
have drawn the Hubble parameter $H(z)$ vs redshift $z$ in figure
26 and the deceleration parameter $q(z)$ vs $z$ in figure 27. We
have seen that the Hubble parameter first decreases and thereafter
slightly increases and deceleration parameter decreases and
crosses the phantom barrier. The function $f(z)$, non-canonical
scalar field $\phi(z)$ and its potential $V(z)$ have been drawn in
figures 28 - 30 respectively. We have seen that $f(z)$ first
increases and then slightly decreases and $\phi(z)$ increase as
$z$ decreases but $V(z)$ first decreases and then sharply
increases as $z$ decreases. Also $f(\phi)$ and $V(\phi)$ in terms
of scalar field $\phi$ have been drawn in figures 31 and 32
respectively. Here $f(\phi)$ sharply increases
but $V(\phi)$ decreases as $\phi$ increases.\\

\section{Discussions and Concluding Remarks}

In this work, we have studied non-canonical scalar field model in
the non-flat $D$-dimensional fractal Universe on the condition
that the matter and scalar field are separately conserved. To get
the solutions of potential $V$, scalar field $\phi$, function $f$,
densities, Hubble parameter and deceleration parameter, the
fractal function has been chosen in the form $v\propto a^{\beta}$.
In the Lagrangian for general form of non-canonical scalar field
model, the kinetic term $\propto f(\phi)X^{n}$ and the function
$f$ has been suitable chosen in the form $f\propto H^{-2n}$. For
$n=2$, the non-canonical scalar field model has been discussed in
ref \cite{Mamon1}. We have chosen four types of parametrizations
forms of equation of state parameter $w_{\phi}(z)$. We have
analyzed best fit values of the unknown parameters ($w_{0},w_{1}$)
of the parametrizations models due to the joint data analysis
(SNIa+BAO+CMB+Hubble). Since we have interested to consider the
$D$-dimensional fractal Universe, so for instance, for graphical
representations, we have assumed $D=5$ which is higher than
4-dimensions and analyzed the physical parameters in this respect.
To get graphical analysis, throughout the paper, we have chosen
all other parameters, $\beta=-0.5,w_{m}=-0.3,\omega=0.2,
v_{0}=0.5,f_{0}=2,n=3,\Omega_{m0}=0.3,\Omega_{k0}=0.05$ and
$H_{0}=72$ km s$^{-1}$ MPc$^{-1}$. For model I, the the best fit
values of the parameters are obtained as
$(w_{0},w_{1})=(-0.738,0.174)$. For model II, the the best fit
values of the parameters are $(w_{0},w_{1})=(-0.796,0.498)$. For
model III, the the best fit values of the parameters are
$(w_{0},w_{1})=(-0.755,0.262)$ and also for model IV, the the best
fit values of the parameters are obtained as
$(w_{0},w_{1})=(-0.765,0.308)$. We have also plotted the contours
of ($w_{0},~w_{1}$) for different confidence levels 66\%, 90\% and
99\% for all these models. For all of the parametrized models, we
have shown that the deceleration parameter $q$ undergoes a smooth
transition from its deceleration phase ($q > 0$) to an
acceleration phase ($q < 0$). For all the models, we have shown in
graphically that the potential function $V(\phi)$ always decreases
and the function $f(\phi)$ always increases as $\phi$ increases.\\

\section*{Acknowledgement} The author UD is thankful to IUCAA,
Pune, India for warm hospitality where part of the work was
carried out. The work of KB was partially supported by the JSPS
KAKENHI Grant Number JP25800136 and Competitive Research Funds for
Fukushima University Faculty (18RI009).\\

\end{document}